\documentclass[accepted]{uai2023} 

\usepackage[british]{babel}

\usepackage{natbib} 
\usepackage{mathtools} 
\usepackage{booktabs} 
\usepackage{tikz} 

\usepackage{hyperref}
\usepackage{natbib}

\usepackage[utf8]{inputenc} 
\usepackage[T1]{fontenc}    
\usepackage{hyperref}       
\usepackage{url}            
\usepackage{booktabs}       
\usepackage{amsfonts}       
\usepackage{nicefrac}      
\usepackage{microtype}     
\usepackage{bm}

\usepackage[makeroom]{cancel}

\usepackage{microtype}
\usepackage{graphicx}
\usepackage{subfigure}
\usepackage{booktabs}

\usepackage{amsbsy}
\usepackage{amsmath}
\usepackage{amsthm}
\usepackage{amsfonts}      
\usepackage{xcolor}
\usepackage{amssymb}
\usepackage{float}
\usepackage{hyperref}
\usepackage{cleveref}

\newtheorem{theorem}{Theorem} 
\newtheorem{corollary}{Corollary}[theorem]
\newtheorem{lemma}[theorem]{Lemma}

\newtheorem{assumption}{Assumption}

\newenvironment{talign*}
 {\csname align*\endcsname}
 {\endalign}
\newenvironment{talign}
 {\csname align\endcsname}
 {\endalign}
\crefname{talign}{}{}
\crefname{equation}{}{}

\usepackage{enumitem}

\usepackage{xcolor}  
\usepackage[linesnumbered,ruled,vlined]{algorithm2e}

\SetCommentSty{mycommfont}
\SetKwInput{KwInput}{Input} 
\SetKwInput{KwOutput}{Output}   

\usepackage[acronym]{glossaries}
\newacronym{IID}{IID}{independent and identically distributed}

\setkeys{glslink}{hyper=false}

\DeclareMathOperator*{\argmin}{arg\,min}

\def\S{\mathcal{S}}

\def\J{\mathcal{J}}

\def\E{\mathbb{E}}

\def\V{\mathbb{V}}
\def\U{\mathcal{U}}

\def\L{\mathcal{L}}

\def\N{\mathbb{N}}

\def\R{\mathbb{R}}
\def\X{\mathcal{X}}
\def\O{\mathcal{O}}

\def\Ttest{T_{\text{test}}}

\def\nepoch{I_{\text{tr}}}

\def\MC{\hat{\E}^{\text{MC}}_{\pi}}
\def\CV{\hat{\E}^{\text{CV}}_{\pi}}

\title{Meta-learning Control Variates: Variance Reduction with Limited Data}

\author[1,3]{\href{mailto:<zhuo.sun.19@ucl.ac.uk>?Subject=Your UAI 2023 paper}{Zhuo Sun}{}}
\author[2,3]{Chris J. Oates}
\author[1,3]{Fran\c{c}ois-Xavier Briol}
\affil[1]{%
    Department of Statistical Science\\
    University College London, London, UK
}
\affil[2]{%
    School of Mathematics, Statistics \& Physics\\
    Newcastle University, UK
  }
\affil[3]{%
    The Alan Turing Institute, London, UK
}

\begin{document}
\maketitle

\begin{abstract}
    Control variates can be a powerful tool to reduce the variance of Monte Carlo estimators, but constructing effective control variates can be challenging when the number of samples is small. 
    In this paper, we show that when a large number of related integrals need to be computed, it is possible to leverage the similarity between these integration tasks to improve performance even when the number of samples per task is very small. Our approach, called \emph{meta learning CVs} (Meta-CVs),  can be used for up to hundreds or thousands of tasks.
    Our empirical assessment indicates that Meta-CVs can lead to significant variance reduction in such settings, and our theoretical analysis establishes general conditions under which Meta-CVs can be successfully trained.
\end{abstract}


\section{Introduction}
Estimating integrals is a significant computational challenge encountered when performing uncertainty quantification in statistics and machine learning. In a Bayesian context, integrals arise in the estimation of posterior moments, marginalisation of hyperparameters, and the computation of predictive distributions. In frequentist statistics, it is often necessary to integrate out latent variables. In machine learning, integrals arise in gradient-based variational inference or reinforcement learning algorithms. 
These problems can usually be formulated as the task of computing
\begin{talign}\label{eq:integral}
   \E_{\pi}[f]  := \int_{\X} f(x) \pi(x) \mathrm{d}x ,
\end{talign}
where $\X \subseteq \R^d$ is the domain of integration, $f:\X \rightarrow \R$ is an integrand, and $\pi:\X \rightarrow [0,\infty)$ is a probability density function. 
(For convenience, we will use $\pi$ to denote both a density and the distribution associated to it.)

It is rare that such integrals can be exactly computed. 
This has led to the development of a range of approximation techniques, including both deterministic and randomised cubature rules.
The focus of this paper is on \emph{Monte Carlo} (MC) methods and their correlated extensions such as \emph{Markov chain Monte Carlo} (MCMC), which make use of a finite collection of evaluations $f(x_i)$ at locations $\{x_i\}_{i=1}^N$ that are randomly sampled; see \citet{Green2015}.

Since the variance of standard MC estimators can be large, control variates (CVs) are often also employed. The idea behind CVs is to approximate $f$ using a suitable family of functions with known integral. Once an approximation $g$ is identified, the CV estimator consist of the sum of $\E_{\pi}[g]$ and a MC (or MCMC) estimator for $\E_{\pi}[f-g]$. An effective control variate is one for which the difference $f-g$ has smaller MC variance than $f$ (or \emph{asymptotic} variance, in the case of MCMC).
CVs have proved successful in a range of challenging tasks in statistical physics \citep{assaraf1999_zerovariance_principleforMCI}, Bayesian statistics \citep{Dellaportas2012, mira2013_zerovariance_MCMC, oates2017_CF_for_MonteCarloIntegration,South2022}, gradient estimation in variational inference \citep{Grathwohl2017,Shi2022} and MCMC \citep{Baker2018}, reinforcement learning \citep{Liu2018,liu2019taming}, and computer graphics \citep{Muller2020}.

Unfortunately, construction of an effective CV usually requires a large number of samples $N$. 
This limits their usefulness in settings when either sampling from $\pi$ or evaluating $f$ is expensive, or when the computational budget is otherwise limited. High-dimensional settings also pose a challenge, since such functions are more difficult to approximate due to the curse of dimensionality. In the latter case, sparsity can be exploited for integrands with low \emph{effective dimension} \citep{South2019,leluc2021_cv_selection_for_montecarlo_integration}, but many integrands do not admit convenient structure that can be easily exploited.

This paper proposes a radically different solution, which \emph{borrows strength} from multiple related integration tasks to aid in the construction of effective CVs. 
Our approach requires a setting where $T$ integration tasks of the form in \eqref{eq:integral} need to be tackled, and where the integrands $f_1,\ldots,f_T$ and densities $\pi_1,\ldots,\pi_T$ are different, but related.
Related integration tasks arise in a broad range of settings, including multifidelity modelling \citep{Peherstorfer2018,Li2022}, sensitivity analysis \citep{Demange-Chryst2022}, policy gradient methods \citep{Liu2018}, and thermodynamic integration \citep{Oates2016thermo}. 
Further examples are considered in \Cref{sec:experiments}, including marginalisation of hyper-parameters in Bayesian inference (see the hierarchical Gaussian process example) and the computation of predictive distributions (see the Lotka--Volterra example). 
In all cases the integrands and densities are closely related, and sharing information across tasks can be expected to deliver a substantial performance improvement.

To date, the only CV method able to exploit related integration tasks is the \emph{vector-valued CVs} of \citet{sun2021_vvCV}. 
This algorithm learns the relationship between integrands through a multi-task learning approach in a vector-valued reproducing kernel Hilbert space. 
It has shown potential, but suffers from a prohibitive computational cost of $\O(T^6)$ and a significant memory cost of $\O(T^2)$. 
The largest experiment in \citet{sun2021_vvCV}, which focused on computation of the model evidence for a dynamical system, was for $T=4$. This lack of scalability in $T$ is a significant limitation; in many of the motivating examples mentioned above, it can be desirable to share information across hundreds or thousands of tasks. 
A key question is therefore: \emph{``How can we construct CVs at scale, sharing information across a large number of tasks?''}

Our answer to this question is an algorithm we call \emph{Meta-learning CVs} (Meta-CVs). As the name indicates, Meta-CVs are built on the meta-learning framework \citep{icml_2017_finn17a_MAML, nips_2018_fin_probabilisticMAML}. The benefits of this approach are three-fold:  (i) the computational cost grows as $\O(T)$, making Meta-CVs feasible for large $T$, (ii) the effective number of parameters for a given task is constant in $T$, limiting significantly the memory cost, and (iii) the construction of the Meta-CV occurs offline, and a new CV can be computed at minimal computational cost whenever a new task arises. 
Before introducing Meta-CVs in \Cref{sec:methods}, we first recall background on CV methods in \Cref{sec:background} and highlighted relevant techniques from related fields in \Cref{sec:related_work}.


\section{Background}
\label{sec:background}
This section contains background information on general techniques used to construct CVs, which will be adapted to Meta-CVs in \Cref{sec:methods}.

\paragraph{Control Variate Methods} 
 In the remainder, we will assume $f$ is in $\L^2(\pi)=\{f:\X \rightarrow \R \text{ s.t. } \E_{\pi}[f^2] < \infty\}$, the space of $\pi$-square-integrable functions on $\X$. 
 This assumption is necessary to ensure the variance of $f$, denoted $\V_{\pi}[f] := \E_{\pi}[f^2]- (\E_{\pi}[f])^2$, exists. 
 The MC estimator of $\E_{\pi}[f]$ is  $\MC[f]= \frac{1}{N}\sum_{i=1}^N f(x_i)$, where $\{x_i\}_{i=1}^N$ are \gls{IID} samples from $\pi$. 
 Under the $\L^2(\pi)$ assumption, this estimator satisfies a central limit theorem:
\begin{talign*}
    \sqrt{N} \left(\MC[f] - \E_{\pi}[f] \right) \rightarrow \mathcal{N}\left(0,\V_{\pi}[f]\right)
\end{talign*}
This result justifies the common use of $\V_{\pi}[f]$ as a proxy for the accuracy of the MC estimator; analogous results hold for MCMC \citep{Dellaportas2012,Belomestny2019, Belomestny2021,Alexopoulos2023} and (randomised) quasi-Monte Carlo \citep{Hickernell2005}, where the \emph{asymptotic variance} and the \emph{Hardy--Krause variation} serve as analogues of $\V_{\pi}[f]$. To limit scope, we focus on MC in the sequel.

A potentially powerful strategy to improve MC estimators is to identify a function $g \in \mathcal{L}^2(\pi)$ for which $\mathbb{V}_\pi[f-g]$ is small, and for which the expectation $\mathbb{E}_\pi[g]$ can be exactly computed. 
From a practical perspective, the identification of a suitable $g$ can be performed using a subset $\{x_i\}_{i=1}^m$ of all available samples (and corresponding integrand evaluations) for the integration task (as described below), and we denote the associated estimator as $\hat{g}_m$.
The selected \emph{control variate} $\hat{g}_m$ forms the basis of an improved estimator
\begin{talign}
    \CV[f] &:= \MC[f- \hat{g}_m] + \E_\pi[\hat{g}_m] \label{eq: CV estimator} \\
    & = \frac{1}{N-m} \sum_{i=m+1}^{N} \left( f(x_i) - \hat{g}_m(x_i) \right)+ \E_\pi[\hat{g}_m] . \nonumber
\end{talign}
Conditional on the training samples $\{x_i\}_{i=1}^m$, a central limit theorem holds for the CV estimator with $\V_{\pi}[f -\hat{g}_m]$ in place of $\V_{\pi}[f]$. 
If $\hat{g}_m$ is an accurate approximation to $f$, the CV estimator will therefore tend to have a smaller error than the original MC estimator. 
Refined analysis is possible when $m$ and $N$ jointly go to infinity and $\hat{g}_m$ converges to a limiting CV, but such asymptotic settings are not representative of the limited data scenarios that motivate this work.

In the remainder of this section, we detail various ways to estimate a CV from data.

\paragraph{Zero-Mean Functions}
A first challenge when selecting a CV is that we require a  \emph{known} mean $\E_\pi[g]$. 
Although \emph{ad-hoc} approaches, such as Taylor expansions of $f$ \citep{paisley2012variational, wang2013variance}, can be used when $\pi$ is relatively simple, this is usually a challenge whenever $\pi$ is a more complex density, such as can be encountered in a Bayesian inference task. 
One way forward is through Stein's method, and we will call any CV constructed in this way a \emph{Stein-based CV}; see \citet{Anastasiou2021}. 
The main components of Stein’s method are a function class $\U$, called a \emph{Stein class}, and an operator $\S_{\pi}$ acting on $\U$, called a \emph{Stein operator}, such that $g:=S_\pi[u]$ satisfies $\E_{\pi}[g]=0$ for any $u\in \U$. One such operator is the \emph{Langevin--Stein operator}
\begin{talign*}
    \S_\pi[u](x) :=  u(x) \cdot \nabla \log \pi(x) +\nabla \cdot u(x)
\end{talign*}
acting on vector fields $u : \X \rightarrow \mathbb{R}^d$. From the divergence theorem, this operator satisfies $\E_\pi[\S_\pi[u]]=0$ under standard tail conditions on the vector field $u$; see \citep{oates2019convergence} for full detail. In addition, evaluation of this operator requires only pointwise evaluation of $\nabla \log\pi(x)$, which is possible even when $\pi$ involves an unknown normalisation constant, i.e. $\pi = \tilde{\pi}/C$ where $\tilde{\pi}$ is known pointwise and $C>0$ is an unknown constant. This is a significant advantage in the present setting since many applications, including problems where $\pi$ is a Bayesian posterior distribution, fall into this category.

The first Stein-based CVs were proposed by \citet{assaraf1999_zerovariance_principleforMCI}, in which $\U$ was a finite-dimensional vector space of functions of the form $u = \nabla p$, with $p$ polynomial of fixed degree; see also \citet{mira2013_zerovariance_MCMC,Papamarkou2014,Friel2014,South2019}.
For additional flexibility, \citet{oates2017_CF_for_MonteCarloIntegration} proposed to take $\U$ to be a Cartesian product of reproducing kernel Hilbert spaces; see also \cite{Oates2016CFQMC,oates2019convergence,Barp2018}, calling this approach \emph{control functionals (CFs)}. 
Since CFs are based on a non-parametric space of functions, they have the capability to approximate complex integrands, but will also have an effective number of parameters growing with $N$, leading to high memory and computational costs. 
It is on these types of CVs that vector-valued CVs are built \citep{sun2021_vvCV}. 
Alternatively, one may take $\U$ to be a (parametric) set of neural networks  \citep{zhu2018neural_CV, Si2020,Ott2023}, or even a combination of neural networks and the aforementioned spaces \citep{south2020_Semi_Exact_CF, Si2020}. 
In this paper, we will focus on CVs constructed with neural networks, which are known as \emph{neural control variates} (Neural-CVs). The rationale for this choice stems from the fact that neural networks are also able to approximate complex functions well, but have a fixed number of parameters, and thus a more manageable memory and computational cost.

\paragraph{Selecting Control Variates} Once a family of CVs has been identified, we need to select from this family an effective CV for the integrand $f$ of interest. We will limit ourselves to parametric families, and will aim to identify a good parameter value so that the variance of the CV estimator is minimised. Let $g(x;\gamma)=\gamma_0 + g_{\gamma_{1:p}}$ where $\gamma := \gamma_{0:p} \in \R^{p+1}$ consists of $p$ parameters $\gamma_{1:p}$ determining the zero-mean Stein-based CV $g_{\gamma_{1:p}}$, and an additional parameter $\gamma_{0}$ that will be used to approximate $\E_{\pi}[f]$. Following the framework of empirical loss minimisation with samples $S = \{x_i, \nabla \log \pi(x_i), f(x_i)\}_{i=1}^m$, the parameter $\gamma$ can be estimated by minimising
\begin{talign*}
    J_S(\gamma) & := \frac{1}{m} \sum_{i=1}^m \left(f(x_i) - g(x_i;\gamma) \right)^2 .
\end{talign*}
The value of $\gamma_0$ minimising this objective is a consistent estimator for $\E_{\pi}[f]$ in the $m \rightarrow \infty$ limit. 
To avoid over-fitting when $m$ is small, penalised objectives have also been proposed \citep{zhu2018neural_CV,Si2020}, but determining the strength of the penalty can represent a very challenging task. 
To limit scope, we proceed to minimise the un-regularised objective in this work.

Conveniently, in the specific case of Neural-CVs, the backward propagation of gradients with respect to the parameters $\gamma$ can be done end-to-end via automatic differentiation techniques implemented in modern deep learning frameworks such as \textsf{PyTorch} \citep{paszke2019pytorch} (which we use in our experiments). Unfortunately, Neural-CVs can require a large number of training samples to learn an accurate approximation to the integrand. 
As a result, Neural-CVs are not well-suited to solving single integration tasks when the total number of samples $N$ is small. The contribution of this work seeks to leverage information from related integration tasks to directly address this weakness of Neural-CVs.

\section{Related Work} 
\label{sec:related_work}

The idea of sharing information across integrations tasks has been explored in a range of settings, each building on a specific structure for the relationship between tasks. Unfortunately, as highlighted below, none of the main approaches can be used in the general setting of large $T$ and arbitrary integrands and densities.

\paragraph{Multi-task Learning for Monte Carlo} \emph{Multi-output Bayesian quadrature} \citep{ICML2018_BQforMultipleRelatedIntegrals,Gessner2019} and vector-valued CVs \citep{sun2021_vvCV} are both approaches based on \emph{multi-task learning}. These methods think of $f_1,\ldots,f_T$ as the output of a vector-valued function with a specific structure shared across outputs, and use this structure to construct an estimator. These approaches can perform very well when the algorithm is able to build on the relationship between tasks, but they also suffer from a computational cost between $\O(T^3)$ and $\O(T^6)$ where $T$ is the number of tasks. These methods are therefore not applicable in settings with a large $T$.

\paragraph{Multilevel and Multi-fidelity Integration} \emph{Multilevel Monte Carlo} \citep{giles2015multilevel} and related methods are applicable in the specific case where $f_1,\ldots,f_T$ are all approximations of some function $f$ with varying levels of accuracy. Although their cost is usually $\O(T)$, these methods are mostly used for problems with small $T$ and where the computational cost of function evaluation varies per integrand. In particular, they are commonly used with a large $N$ for cheaper but less accurate integrands, and a small $N$ for expensive but accurate integrands. This setting is therefore different from that considered in the present work.

\paragraph{Monte Carlo Methods for Parametric and Conditional Expectations}
\emph{Parametric expectation} or \emph{conditional expectation} methods \citep{Longstaff2001,krumscheid2018multilevel} consider the task of approximating $\E_{X \sim \pi}[f(X,y)]$ or $\E_{X \sim \pi(\cdot|Y=y)}[f(X)]$ uniformly over $y$ in some interval. These methods can be applied when $T$ is large, but they usually rely on a specific structure of the problem: smoothness of these quantities as $y$ varies. The methodological development in our work does not rely on smoothness assumptions of this kind.

\paragraph{Importance Sampling} Importance sampling is commonly used to tackle an integration task with respect to $\pi$ when samples from a related distribution $\pi'$ are available. It works by weighting samples according to the ratio $\pi/\pi'$, and is applicable to multiple tasks with a $\O(T)$ cost. However, the challenge is that $\pi'$ needs to be chosen carefully in order for the estimator to have low variance. The problem of multiple related integrals was considered by \cite[Section 8]{Glynn1989} and \cite{Madras1999,Demange-Chryst2022}, where the authors seek an importance distribution $\pi'$ which performs well across a range of tasks. However, identifying such an importance distribution will usually not be possible when $T$ is large.

\section{Methodology}
\label{sec:methods}

We now set out the details of our proposed Meta-CVs.

\paragraph{Problem Set-up}
 Consider a finite (but possibly large) number, $T$, of integration tasks 
 \begin{talign*}
     \E_{\pi_1}[f_1],\ldots, \E_{\pi_T}[f_T]
 \end{talign*}
and denote by $\mathcal{T}_t:= \{f_t, \pi_t\}$ the components of the  $t^{\text{th}}$ task, consisting of a density $\pi_t:\X \rightarrow [0,\infty)$ and an integrand $f_t \in \mathcal{L}^2(\pi_t)$. For each task, we assume we have access to data of the form
\begin{talign*}
    D_t = \{x_i,\nabla \log\pi_t(x_i),f_t(x_i)\}_{i=1}^{N_t},
\end{talign*}
where $N_t \in \N^+$ is relatively small. 
In addition, we will assume that these tasks are related.
Informally, we may suppose that $\mathcal{T}_1,\ldots,\mathcal{T}_T$ are independent realisations from a distribution over tasks arising from an \emph{environment}, but we do not attempt to make this notion formal. 
This set-up allows us to frame Meta-CVs in the framework of gradient-based meta-learning.

\paragraph{Meta-learning CVs} 
\label{sec:mcv_construction}

\emph{Gradient-based meta learning} \citep{icml_2017_finn17a_MAML, finn2019online_maml, iclr_2018_grant_recasting_HBMAML, NIPS_2018_yoon_BayesianMAML, sun2021amortized} was first proposed in the context of \emph{model-agnostic meta-learning} \citep{icml_2017_finn17a_MAML, finn2019online_maml}. 
It was originally designed for ``learning-to-learn'' in a supervised-learning context, with a specific focus on regression and image classification. The focus of this approach is on the ability to rapidly adapt to new tasks. This is achieved by identifying a meta-model, which acts as an initial model which can be quickly adapted to a new task by taking a few steps of some gradient-based optimiser on its parameters.

In this paper we adapt gradient-based meta learning to the construction of CVs. This leads to a two-step approach: 
The first step, highlighted in Algorithm \ref{alg:meta_neural_cv_train}, consists of learning a \emph{Meta-CV}, a CV that performs  ``reasonably well for most tasks''.
The second step, highlighted in Algorithm \ref{alg:meta_neural_cv_test}, consists of fine-tuning this Meta-CV to each specific task, using a few additional steps of stochastic optimisation on a task-specific objective function, to obtain a \emph{task-specific} CV.

Before describing these algorithms, for each task $\mathcal{T}_t:=\{f_t, \pi_t\}$, we split the samples into two disjoint sets $D_t = S_t \cup Q_t$, so that
\begin{talign*}
    S_t & :=\{x_j, \nabla \log \pi_t(x_j), f_t(x_j)\}_{j=1}^{m_t}\\
    Q_t & :=\{x_j, \nabla \log \pi_t(x_j), f_t(x_j)\}_{j=m_t+1}^{N_t}.
\end{talign*}
The roles of these two datasets will differ depending on whether the task is used for training the Meta-CV, or for deriving a task-specific CV, and we will return to this point below. For simplicity, all of our experiments will consider $m_t = N_t/2$. 
Note that these datasets correspond to the concepts of the \emph{support set} and the \emph{query set} in the terminology of gradient-based meta learning \citep{icml_2017_finn17a_MAML, finn2019online_maml}. 


\paragraph{Constructing the Meta-CV}
The first step in our method is to construct a Meta-CV; this will later be fine-tuned into a task-specific CV. Here we will follow the approach in \Cref{sec:background} and use a flexibly-parametrised Neural-CV.

To decouple the choice of optimisation method from the general construction of a Meta-CV,  $L$ steps of an arbitrary gradient-based optimiser will be denoted $\textsc{Update}_L(\gamma, \nabla_{\gamma} J\left(\gamma\right);\alpha)$, where $\gamma \in \mathbb{R}^{p+1}$ is the initial parameter value, $\nabla_{\gamma} J\left(\gamma\right)$ is the gradient of an objective $J: \mathbb{R}^{p+1} \rightarrow \R$, and $\alpha$ represents  parameters of the optimisation method. Popular optimisers include gradient descent and Adam \citep{Kingma2015}, but more flexible alternatives also exist \citep{andrychowicz2016_l2l_GD_by_GD,grefenstette2019generalized_maml}. 
For example, the update corresponding to $L$-step gradient descent starting at $\gamma_0$ consists of  $\gamma_j := \gamma_{j-1} - \alpha \nabla J\left(\gamma_{j-1} \right)$ for $j=1,\ldots, L$ . 
Using this notation, we can represent an \emph{idealised Meta-CV} as a CV whose parameters satisfy
\begin{talign}
\label{eq:mcv_obj}
  \gamma_{\text{meta}}  & \in \argmin_{\gamma \in \mathbb{R}^{p+1}} \E_{t} \left[ J_{t} \left( \textsc{Update}_L \left(\gamma, \nabla_{\gamma} J_{t}\left(\gamma \right);\alpha \right)\right) \right] , 
\end{talign}
where $\mathbb{E}_t$ denotes expectation with respect to a uniformly sampled task index $t \in \{1,\dots,T\}$. This objective is challenging to approximate since it requires solving nested optimisation problems. We therefore follow the approach in \cite{icml_2017_finn17a_MAML} and use a gradient-based bi-level optimisation scheme described in Algorithm \ref{alg:meta_neural_cv_train}. 
This requires estimating the gradient of the loss $J_{t}$ in both the inner and outer level. 
To prevent over-fitting, we do this using two independent datasets: $S_t$ and $Q_t$. 
We will call the output of Algorithm \ref{alg:meta_neural_cv_train}, denoted $\hat{\gamma}_{\text{meta}}$, our \emph{meta-parameter}, and $g(\cdot;\hat{\gamma}_{\text{meta}})$ will be called the Meta-CV.

\begin{algorithm}
\caption{Learning a Meta-CV}
\label{alg:meta_neural_cv_train} 
  
  \KwInput{Training tasks $\mathcal{T}_1,\ldots,\mathcal{T}_{T}$, initial parameter $\gamma_0$, $\textsc{Update}$ rule, $\#$ update steps $L$, optimiser parameters $\alpha$ and $\eta_1,\ldots,\eta_{\nepoch}$, mini-batch size $B$, $\#$ meta-iterations $\nepoch$.  }
    
        \For{ $i =  1,\dots,\nepoch$}    
        { 
        Sample $t_1,\dots,t_B$ uniformly from $\{1,\dots,T\}$.

          \For{$t \in \{t_1,\ldots,t_B\}$}
          {
              Initialize $\gamma^t_0 \leftarrow \gamma_{i-1}$.

             \For{$j = 1,\dots,L$}
              {     
               \small $\gamma^t_{j} \leftarrow  \textsc{Update}(\gamma_{j-1}^t,  \nabla_{\gamma_{j-1}^t}             J_{S_t}(\gamma_{j-1}^t) ; \alpha  )$.
              }
          }
         
          \small $\gamma_{i} \leftarrow  \textsc{Update}(\gamma_{i-1} ,  \frac{1}{B} \sum_{b=1}^{B} \nabla_{\gamma_{i-1}} J_{Q_{t_b}}(\gamma_L^{t_b}) ; \eta_i )$. 
        }
\KwOutput{The meta-parameter $\hat{\gamma}_{\text{meta}} := \gamma_{\nepoch}$.}
\end{algorithm}

\paragraph{Task-Specific CVs} Once a meta-parameter $\hat{\gamma}_{\text{meta}}$ has been identified,
for each task $\mathcal{T}_{t}$ we only need to adapt the meta-parameter through a few optimisation steps to obtain a task-specific parameter, $\hat{\gamma}_t$, and hence a corresponding task-specific CV $g(\cdot;\hat{\gamma}_t)$. This can be done by using Algorithm \ref{alg:meta_neural_cv_test}, and can be applied either to one of the $T$ tasks in the training set, or indeed to an as yet unseen task. 
Once such a task-specific CV is identified, we can simply use the CV estimator in Equation \eqref{eq: CV estimator} to estimate the corresponding integral $\E_{\pi_t}[f_t]$. Note that we once again use two datasets per task, but their role differs from that in Algorithm \ref{alg:meta_neural_cv_train}: $S_t$ will be used for selecting the task-specific CV $g(\cdot;\hat{\gamma}_t)$ through Algorithm \ref{alg:meta_neural_cv_test}, whilst $Q_t$ will be used to evaluate the CV estimator in \eqref{eq: CV estimator}.

To understand how these task-specific CVs \emph{borrow strength}, we highlight that the task-specific CV are constructed using $\sum_{t=1}^{T} N_{t}$ samples in total. Thus, when $T$ is large and $N_t$ is small, our  task-specific CV may be based on a much larger number of samples compared to any CV constructed solely using data on a single task. The closeness of the relationship between tasks of course determines the value of including these additional data into the training procedure for a CV; this will be experimentally assessed in \Cref{sec:experiments}.

\begin{algorithm}
\caption{Task-specific CVs from the Meta-CV}
\label{alg:meta_neural_cv_test} 
  \KwInput{Integration task $\mathcal{T}_{t}$, meta-parameter $\hat{\gamma}_{\text{meta}}$, $\textsc{Update}$ rule, $\#$ update steps $L$, optimiser parameters $\alpha$.}

              Initialize $\gamma_0 \leftarrow \hat{\gamma}_{\text{meta}}$.
              
            \For{$j = 1,\dots,L$}
              { 
               $\gamma_{j} \leftarrow  \textsc{Update}(\gamma_{j-1} , \nabla J_{S_{t}}(\gamma_{j-1}) ;\alpha) $.
               }

  \KwOutput{Task-specific parameter $\hat{\gamma}_t:=\gamma_L$.}
          
\end{algorithm}

\paragraph{Computational Complexity}  
To discuss the complexity of our method, suppose first that the parameter $\hat{\gamma}_{\text{meta}}$ of the Meta-CV has already been computed.
The additional computational complexity of training all task-specific Neural-CVs is then $\O(T L)$, where $L$ is the number of optimisation steps used to fine-tune the CV to each specific task.
Ordinarily a large number of optimisation steps are required to learn parameters of a neural network, but due to meta-learning we expect the number $L$ of steps required to fine-tune task-specific CVs to be very small (indeed, we take $L=1$ in most of our experiments). This is because as $L$ grows, the task-specific CV is less and less dependent on the meta-CV; see also \cite{antoniou2018train}.
In addition, taking $L$ to be small means fine-tuning a Meta-CV can be orders of magnitude faster compared to training Neural-CVs independently for each task.

Of course, we also need to consider the complexity of training the Meta-CV.
This can require a large number $L$ of optimisation steps in general - a point we assess experimentally in \Cref{sec:experiments} and theoretically in \Cref{sec:theory} - but this number is broadly comparable to that required to train a Neural CV to a single task. However, the scaling in $p$, the number of neural network parameters, is at least $\O(p^2)$ \citep{fallah2020convergence_gbml} for our approach (due to second-order derivatives in Algorithm \ref{alg:meta_neural_cv_train}) against $\O(p)$ for Neural-CVs. This will be a challenge for our method when $p$ is large, and we will return to this issue in the conclusion of the paper.


\section{Experimental Assessment}
\label{sec:experiments}
The performance of the proposed Meta-CV method will now be experimentally assessed, using a range of problems of increasing complexity where $N_t$ is small and $T$ is large. 
For simplicity, we will limit ourselves to the setting where the $N_t$ are equal and where the Adam optimiser is used. 
The existing methods discussed in \Cref{sec:related_work} \emph{cannot be applied to the problems in this section} due to the large value of $T$ and associated prohibitive computational cost, and we therefore only compare to methods which do not borrow strength between tasks: MC, Neural-CVs and CFs.
The code to reproduce our results is available at: \url{https://github.com/jz-fun/Meta_Control_Variates}.

\paragraph{A Synthetic Example}
\label{sec:experiments-oscillatory}
Trigonometric functions are common benchmarks for meta-learning \citep{icml_2017_finn17a_MAML, iclr_2018_grant_recasting_HBMAML} and CVs \citep{oates2017_CF_for_MonteCarloIntegration, oates2019convergence}. 
Consider integrands of the form 
\begin{talign*}
f_t(x;a_t) = \cos\left( 2 \pi a_{t,1} + \sum_{i=1}^d  a_{t,i+1} x_i \right), 
\end{talign*}
with parameters $a_t \in \R^{d+1}$, 
and let $\pi_t$ be the uniform distribution on $\X = [0,1]^d$.
The integrals $\mathbb{E}_{\pi_t}[f_t]$ can then be explicitly computed and serve as a ground truth for the purpose of assessment. Note that $a_t$ controls the difficulty of the $t^\text{th}$ integration task.
To generate related tasks, we sample the $a_t$ from a distribution $\rho$ consisting of independent uniforms; see \Cref{appdx:experiments_oscillatory} for full detail. 

\begin{figure}[t!]
    \centering
    \includegraphics[width=1.\columnwidth, trim={0.2cm 0.cm 0.cm 0.cm},clip]{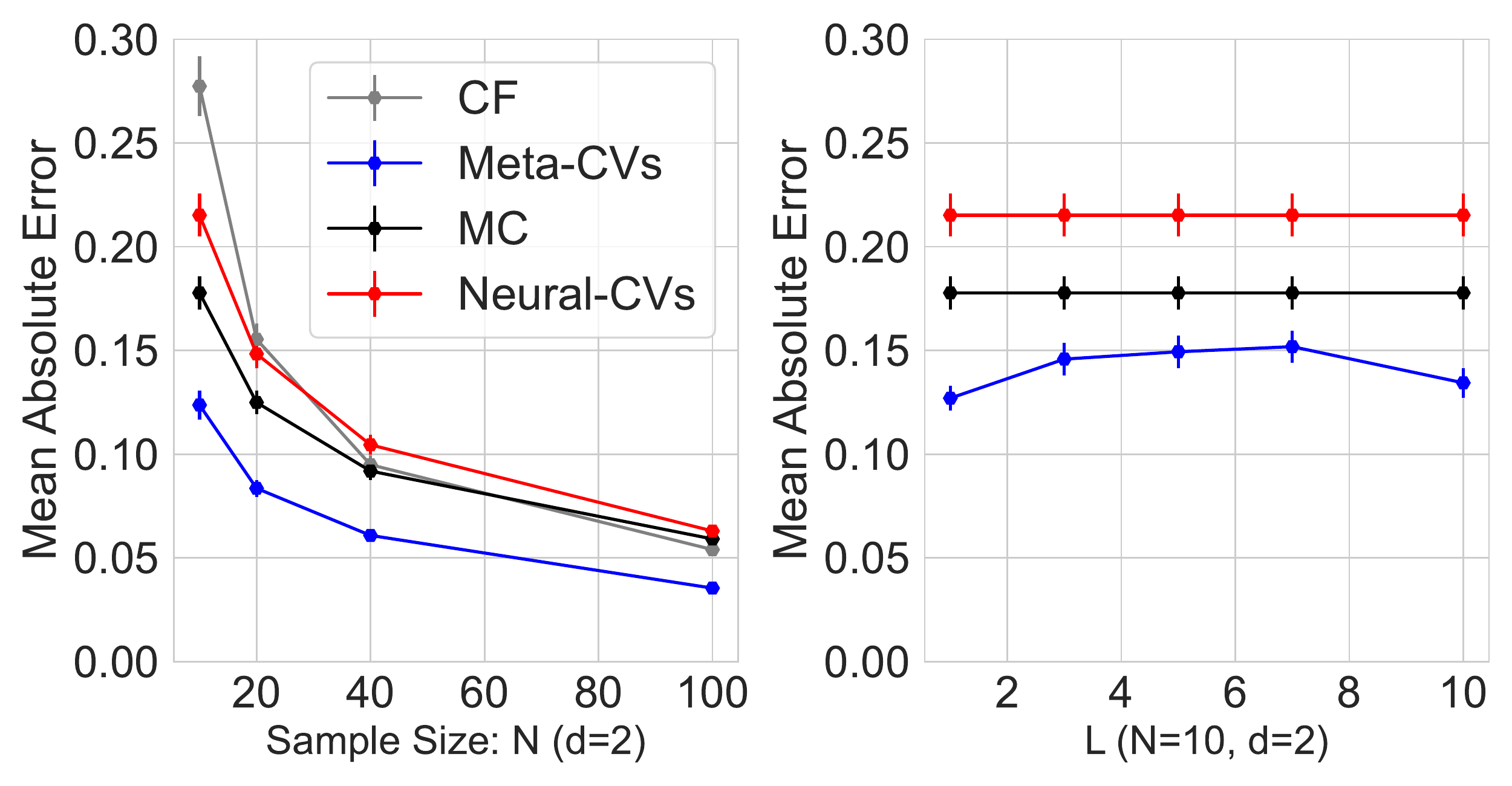}
    \caption{Mean absolute error (with $95\%$ confidence intervals) for $\Ttest=1,000$ oscillatory functions (with $N_t=N$ and $m_t=n_t=\nicefrac{N}{2}$ for all $t$). \emph{Left}: Increasing sample size $N_t$ when $d=2$ (Meta-CVs with $L=1$); \emph{Right}: Increasing number of inner gradient steps $L$ of Meta-CVs.
    }
    \label{fig:oscillatory_invest_insteps_N_and_m}
\end{figure}

\begin{figure}[t!]
    \centering
    \includegraphics[width=1.\columnwidth, trim={0.2cm 0.2cm 0.2cm 0.cm},clip]{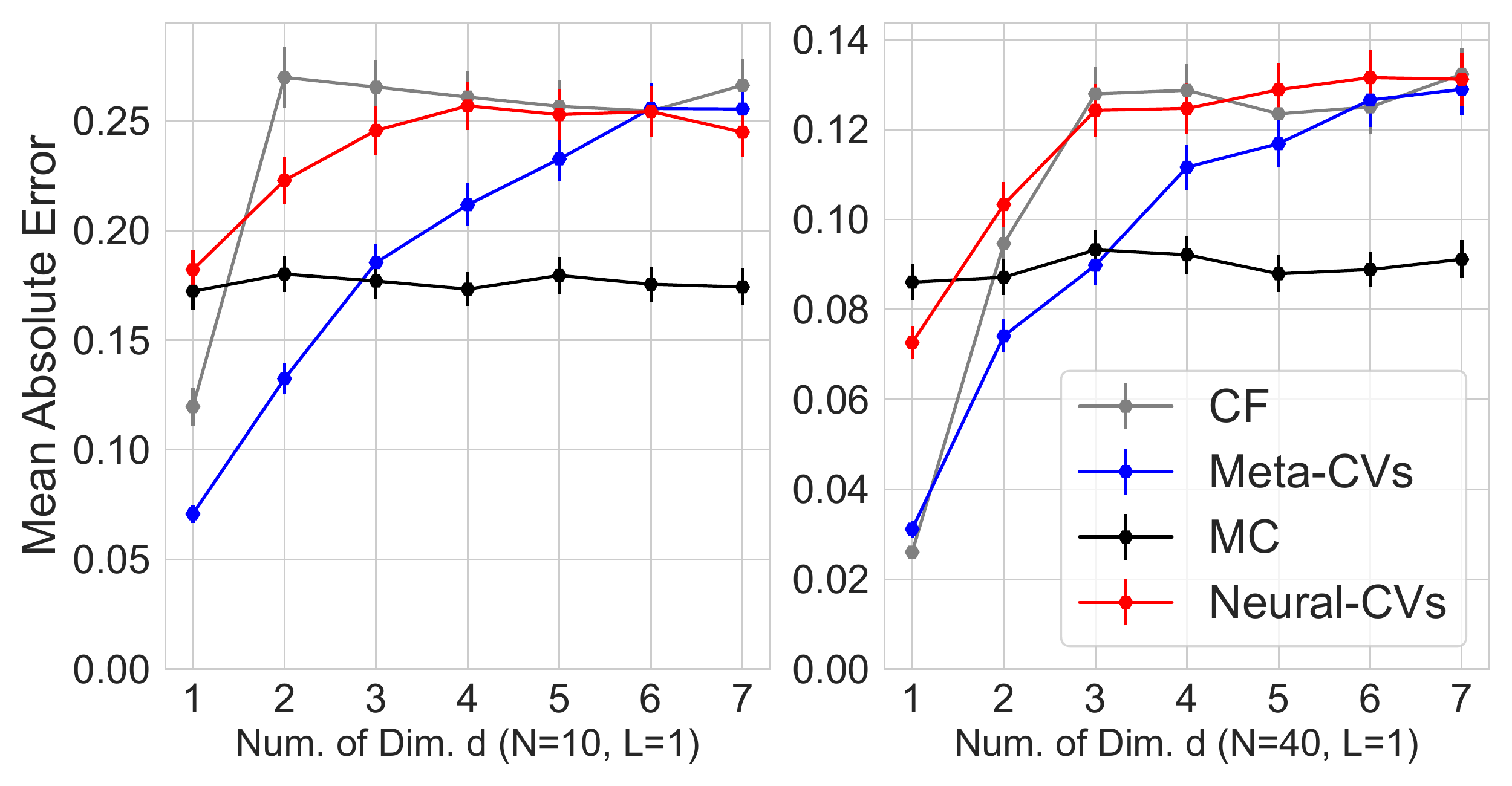}
    \caption{Mean absolute error (with $95\%$ confidence intervals) for $\Ttest=1,000$ oscillatory functions for increasing dimension $d$ (with $N_t =N$ and $m_t = n_t = \nicefrac{N}{2}$ for all $t$).
    }
    \label{fig:oscillatory_dim_invest}
\end{figure}

\begin{figure}[t!]
    \centering
    \includegraphics[width=0.7\columnwidth, trim={0.2cm 0.2cm 0.2cm 0.cm},clip]{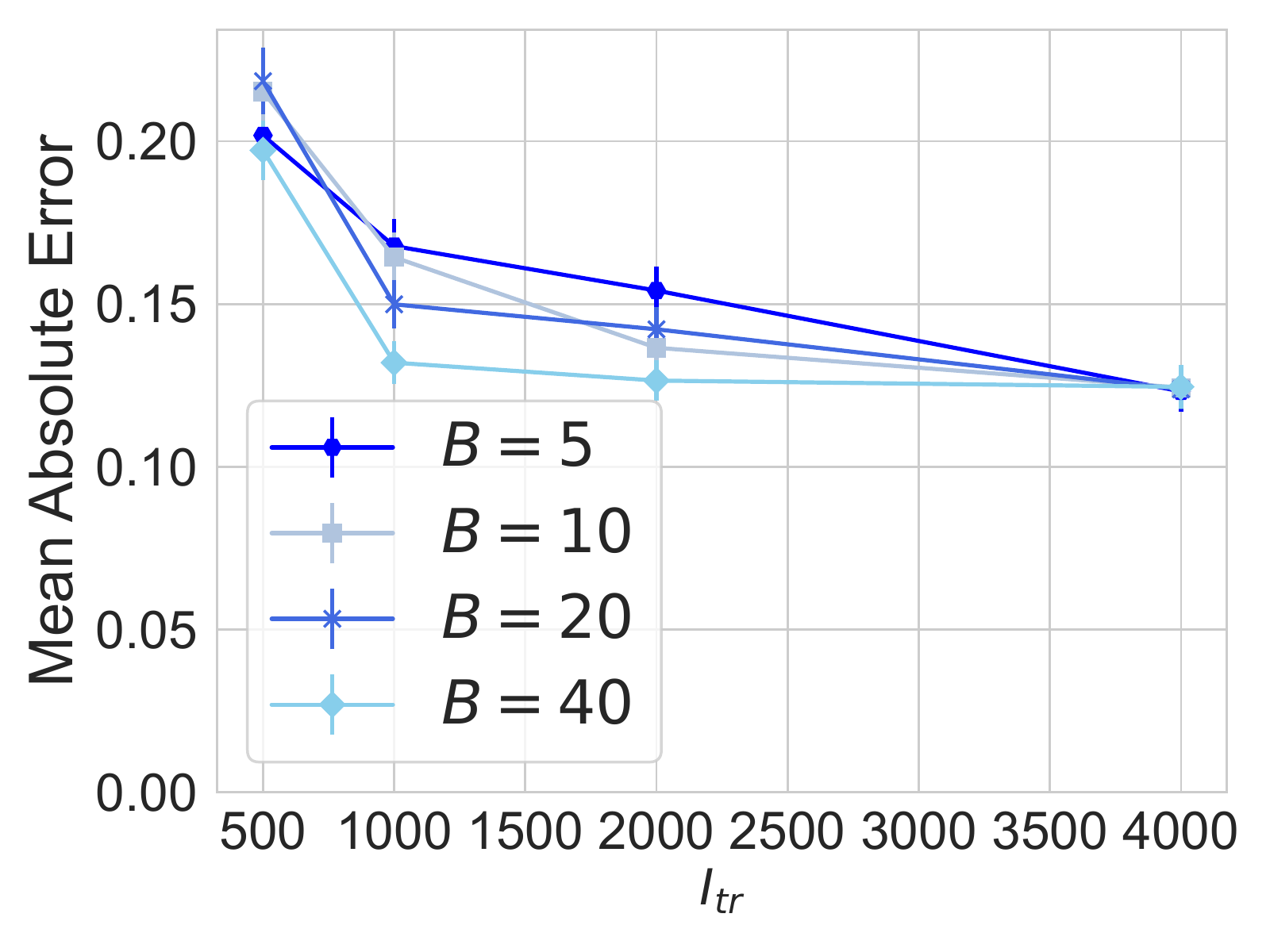}
    \caption{Mean absolute error (with $95\%$ confidence intervals) of Meta-CVs for $\Ttest=1,000$ $2$-dimensional oscillatory functions for increasing $B$ and $\nepoch$ (with $L=1$, $N_t = N$ and $m_t = n_t = \nicefrac{N}{2} = 5$ for all $t$).
    }
    \label{fig:oscillatort_invesr_B_Itr}
\end{figure}

\Cref{fig:oscillatory_invest_insteps_N_and_m} considers the case $d=2$, where we train the Meta-CVs on $T =20,000$ tasks in total. 
To challenge Meta-CVs, all methods were assessed in terms of their performance evaluated on an \emph{additional} $\Ttest=1,000$ tasks, not available during training of the Meta-CV.
On the left panel, we consider the performance of CVs as we increase the sample size $N_t$ per task. Regardless of the sample size considered, Meta-CVs outperform MC, CFs, and Neural-CVs in terms of mean absolute error over new unseen tasks. 
This can be explained by the fact that Meta-CVs is the only method which can transfer information across tasks, able to exploit the large training dataset. 
In this example, Neural-CVs and CF perform even worse than MC when $N_t$ is small, highlighting the challenge of using CVs in these settings.  In the right panel of \Cref{fig:oscillatory_invest_insteps_N_and_m}, we investigate the effect of the number of gradient-based updates $L$, which shows the robustness of Meta-CVs to $L$.  
We investigate the performance of these methods as $d$ increases in \Cref{fig:oscillatory_dim_invest}. Clearly, all CVs suffer from a curse of dimensionality, but Meta-CVs do improve on the other CVs for $d <6$. Alleviating this curse of dimensionality could be an important direction for future research in CVs. We also investigate the effect of $B$ and $\nepoch$ in \Cref{fig:oscillatort_invesr_B_Itr} by comparing the resulting performance of Meta-CVs on $1000$ $2$-dimensional unseen test tasks. It is found empirically that a larger value of $B$ helps to achieve the optimal performance faster; and a large value of $\nepoch$ results in improvement in performance as expected.

We conclude with a brief discussion of computational cost. The cost of computing independent Neural-CVs on all unseen test tasks is around $2$ minutes. In contrast, the offline time for training the Meta-CV is around $7$ minutes (with $L=1$ when $B=5$, $\nepoch = 4000$, $N_t=10$ and $d=1$), but the online time taken for deriving task-specific CVs for all the same $1000$ unseen test tasks is approximately $6$ seconds in total. This demonstrates that our Meta-CV can be rapidly adapted to new tasks.

\paragraph{Uncertainty Quantification for Boundary Value ODEs}
Our second example considers the computation of expectations of functionals of physical models represented through differential equations. The expectations are taken with respect to expert-specified distributions over parameters of these models, with the aim of performing uncertainty quantification. We consider a boundary-value ODE with unknown forcing closely resembling that of \citet{giles2015multilevel}:
\begin{talign*}
    \frac{\mathrm{d}}{\mathrm{d}s} ( c(s) \frac{\mathrm{d}u}{\mathrm{d}s}) = -50 x^2, \qquad  0<s<1,
\end{talign*}
with boundary $u(0)=u(1)=0$, $c(s) = 1+a s$. The integrand of interest is $f_t(x) = \int_{0}^1 u(s,x;a_t)\mathrm{d}s$ where $a_t$ are draws from $\rho = \textsf{Unif}(0,1)$, and the integral of interest is $\E_{X \sim \pi_t}[f_t(X)]$ where each $\pi_t = \mathcal{N}(0,1)$.  We use a finite difference approximation of $f_t$ described in \citet{giles2015multilevel}; see \Cref{appdx:boundary_value_ODEs} for detail. This is a relatively simple example, but it is representative of a broader class of challenging problems where improved numerical methods are needed to approximate integrals due to a large cost per integrand evaluation and therefore limited $N_t$. 

\begin{figure}[t!]
    \centering
    \includegraphics[width=1.\columnwidth, trim={0.2cm 0.cm 0.2cm 0.cm},clip]{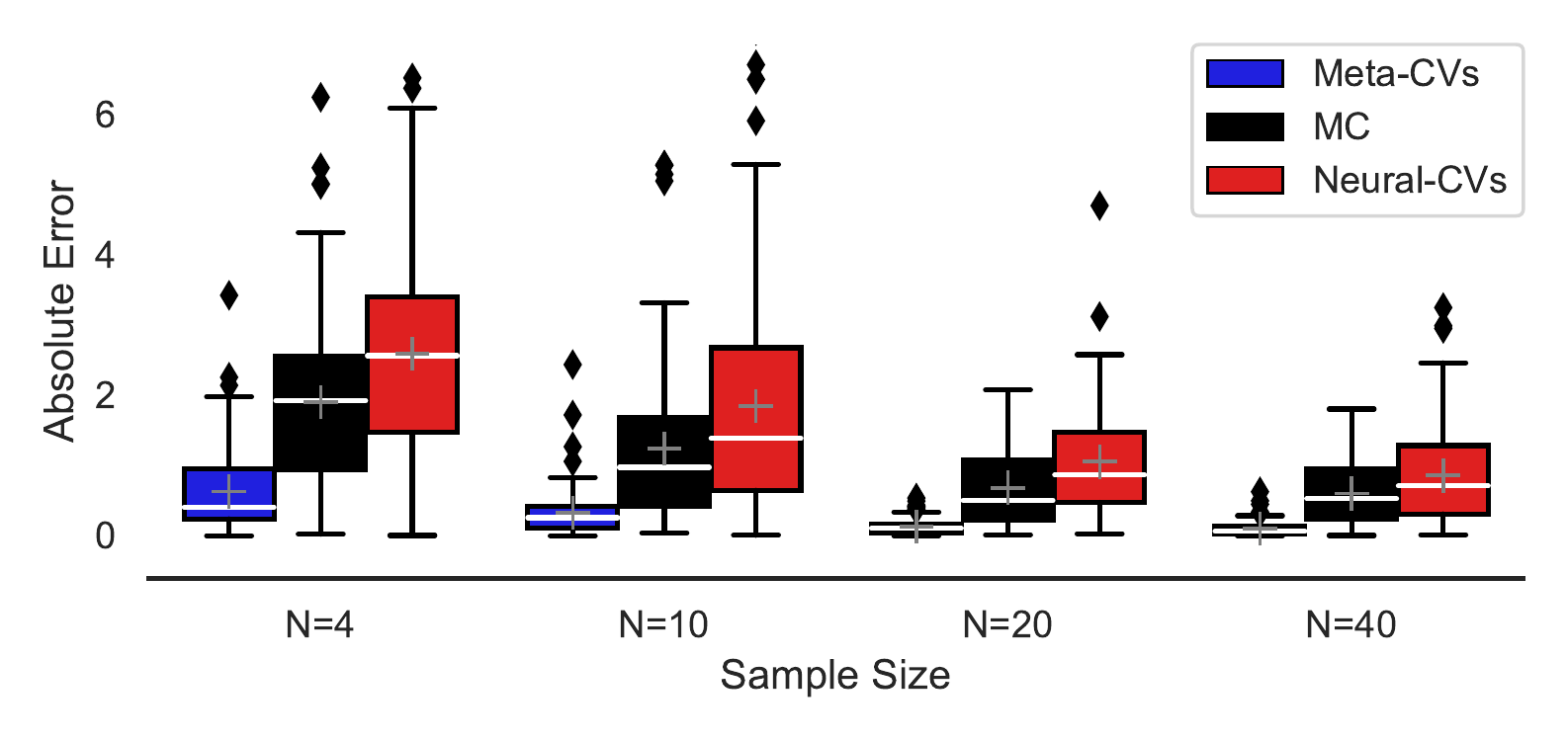}
    \caption{Absolute error for $\Ttest = 100$ (with $N_t = N$ and $m_t = n_t =\nicefrac{N}{2}$ for all $t$.) unseen tasks from the boundary value ODE problem.} 
    \label{fig:bound_value_ode}
\end{figure}

The results are presented in \Cref{fig:bound_value_ode}. We compare the performance of Meta-CVs with MC and Neural-CVs on $\Ttest = 100$ unseen tasks  (grey crosses are mean absolute errors; white horizontal lines are medians). For this example, Meta-CVs outperform Neural-CVs and MC consistently in all cases, highlighting once again the benefits of sharing information across a large number of tasks when $N_t$ is  small.

\paragraph{Bayesian Inference for the Lotka--Volterra System}
Our next example also considers uncertainty quantification for differential equation-based models, but this time in a fully Bayesian framework. In particular, we consider a parametric ODE system, the Lotka--Volterra model \citep{lotka1927fluctuations}, commonly used in ecology and epidemiology, given by
\begin{talign*}
    \frac{\mathrm{d}u_1}{\mathrm{d}s} = x_1 u_1- x_2 u_1 u_2 , \; \frac{\mathrm{d}u_2}{\mathrm{d}s} = x_3 u_1 u_2- x_4 u_2, 
\end{talign*}
where $u_1(s)$ and $u_2(s)$ are the numbers of preys and predators at time $s$, and $u_1(0)=x_5$ and $u_2(0)=x_6$. 
Suppose we have access to observations of $u = (u_1,u_2)$ at time points $\{s_1, \ldots s_q\}$, corrupted with independent log-normal noise with variances $x_7$ and $x_8$ respectively. 
A `task' here corresponds to computing the posterior expectation of model parameters $x$ for a given dataset; different datasets, which could for example correspond to different animal species, or to different geographical regions, determine the posterior distribution $\pi_t$ of interest. Bayesian inference on this type of ecological \citep{bolker2008ecological} and epidemiological \citep{brauer2017mathematical} models is challenging due to the high cost of MCMC sampling, significantly limiting the number of effectively independent samples $N_t$. In our experiments, we use the dataset from \citet{hewitt1921conservation} on snowshoe hares (preys) and Canadian lynxes (predators). 
We sub-sample the whole dataset to mimic the process of sampling sub-populations and our goal is to learn a Meta-CV which can be quickly adapted to new sub-populations observed in the future; see \Cref{appdx:experiments_lotka} for full detail.

\begin{figure}[t!]
    \centering
   \includegraphics[width=\columnwidth, trim={0.1cm 0.2cm 0.2cm, 0.1cm},clip]{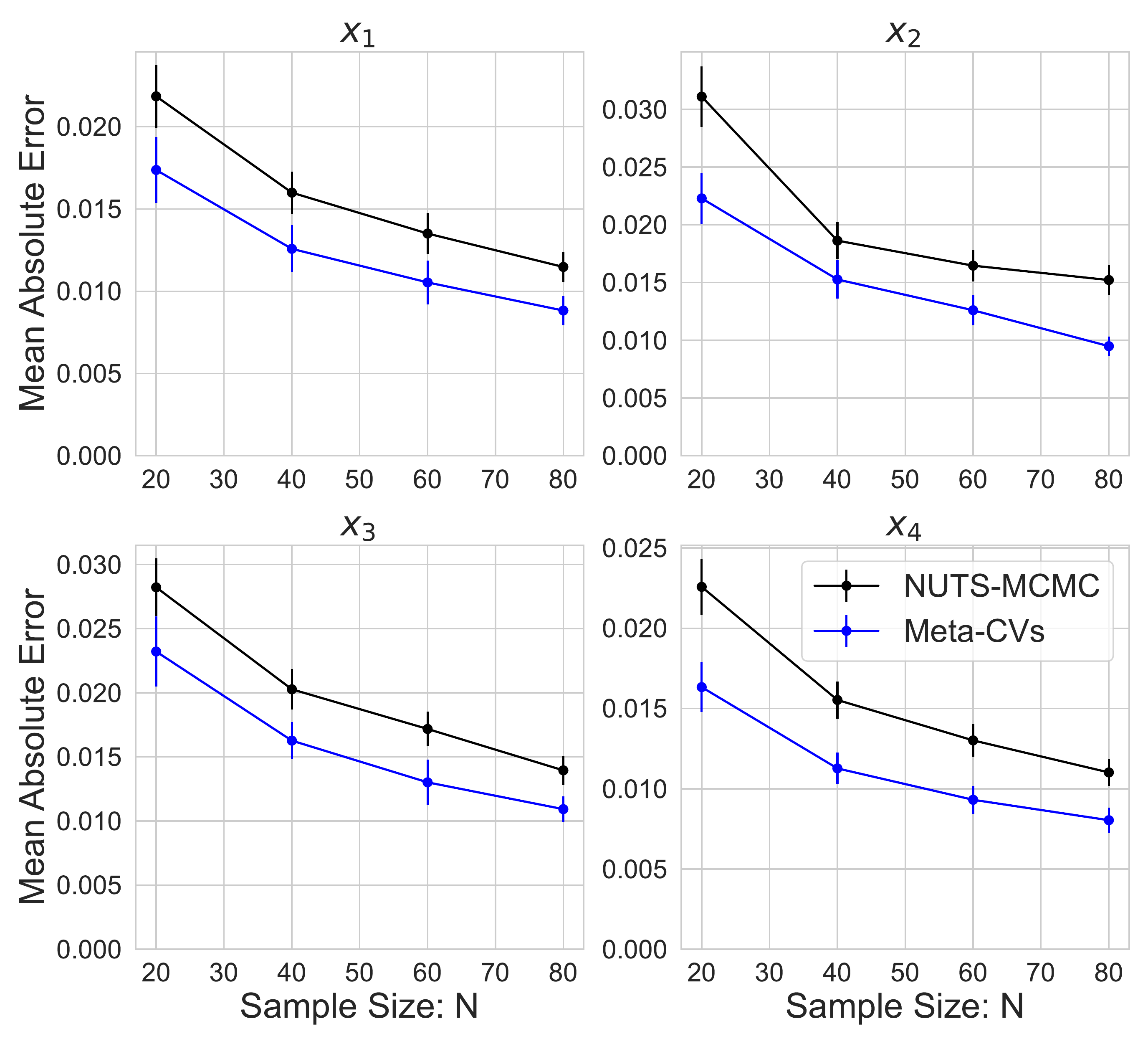}
    \caption{Mean absolute errors (with $95\%$ confidence intervals) over $40$ sub-populations for varying $N_t$. Here, $N_t=N$ and $m_t = n_t = \nicefrac{N}{2}$ for all $t$.}
    \label{fig:lotka}
\end{figure}

Results are presented in \Cref{fig:lotka}. We compare Meta-CV to MCMC (a \emph{No-U-Turn Sampler} (NUTS) implemented in Stan \citep{carpenter2017stan}).  As previously discussed, CVs perform poorly in high-dimensions when $N_t$ is small. This is exactly what we observe: Neural-CVs performs between $5-$ and $12-$times worse than MCMC and is therefore not included in the figure. In contrast, Meta-CVs is able to achieve a lower mean absolute error than MCMC for the values of $N_t$ considered, demonstrating the clear advantage of sharing information across tasks for higher-dimensional problems.

\paragraph{Marginalization in Hierarchical Gaussian Processes}
Marginalisation of hyper-parameters is a common problem in Bayesian statistics. We consider a canonical example for hierarchical Gaussian process regression \citep{rasmussen2003_GPML}, which was tackled with CVs by \citet{oates2017_CF_for_MonteCarloIntegration}. 
The problem consists of recovering an unknown function $\nu$ describing a $7$ degrees-of-freedom Sarcos anthropomorphic robot arm, from a $21$-dimensional input space, based on a subset of the dataset described in \citet{rasmussen2003_GPML}. 
Data consist of observations $y_i = \nu(z_i)+\epsilon_i$ at inputs $z_i$ for $i =1,\ldots, q$, where $\epsilon_i$ are \gls{IID} zero-mean Gaussian random variables with known standard deviation $\sigma > 0$. 
A zero-mean Gaussian process prior is placed on $\nu$, with covariance function $k_{x}(z,z') = x_1 \exp(-\|z-z'\|_2^2/2x^2_2)$, as well as priors on the hyper-parameters $x = (x_1,x_2)$.  
Given observations $y_{1:q} = (y_1,\ldots,y_q)^\top$, we consider the `task' of predicting the response $\nu(x^*)$ at an unseen state $z^*$, marginalising out any posterior uncertainty associated with the hyperparameters $x$ of the Gaussian process model. 
This can be achieved through the Bayesian posterior predictive mean $\E[Y^*|y_{1:q}] = \E_{X \sim \pi(\cdot|y_{1:q})}[ \E[Y^*|y_{1:q}, X]]$. 
This is an integral of 
\begin{align*}
f(x) &= \E[Y^*|y_{1:q}, x] \\
&=  K_{z^*, q}(x)  (K_{q, q}(x)+\sigma^2 I_q)^{-1} y_{1:q} 
\end{align*}
against the posterior on hyperparameters $\pi(x|y_{1:q})$, where  $(K_{q,q}(x))_{i,j}= k_{x}(z_i,z_j)$ and $(K_{z^*,q}(x))_{j}= k_{x}(z^*, z_j)$ for $i, j \in \{1, \ldots, q\}$. 
The integrand is therefore an expensive function: $\O(q^3)$ operations are needed per evaluation, which will be significant when $q$ is beyond a few hundred. However, it is also common to want to compute this quantity for several new inputs $z_1^*,\ldots,z_T^*$, leading to closely related integrands $f_1,\ldots,f_T$ whose relationship could potentially be leveraged.

\begin{figure}[H]
    \centering
   \includegraphics[width=.95\columnwidth, trim={0.2cm 0.15cm 0.1cm 0cm},clip]{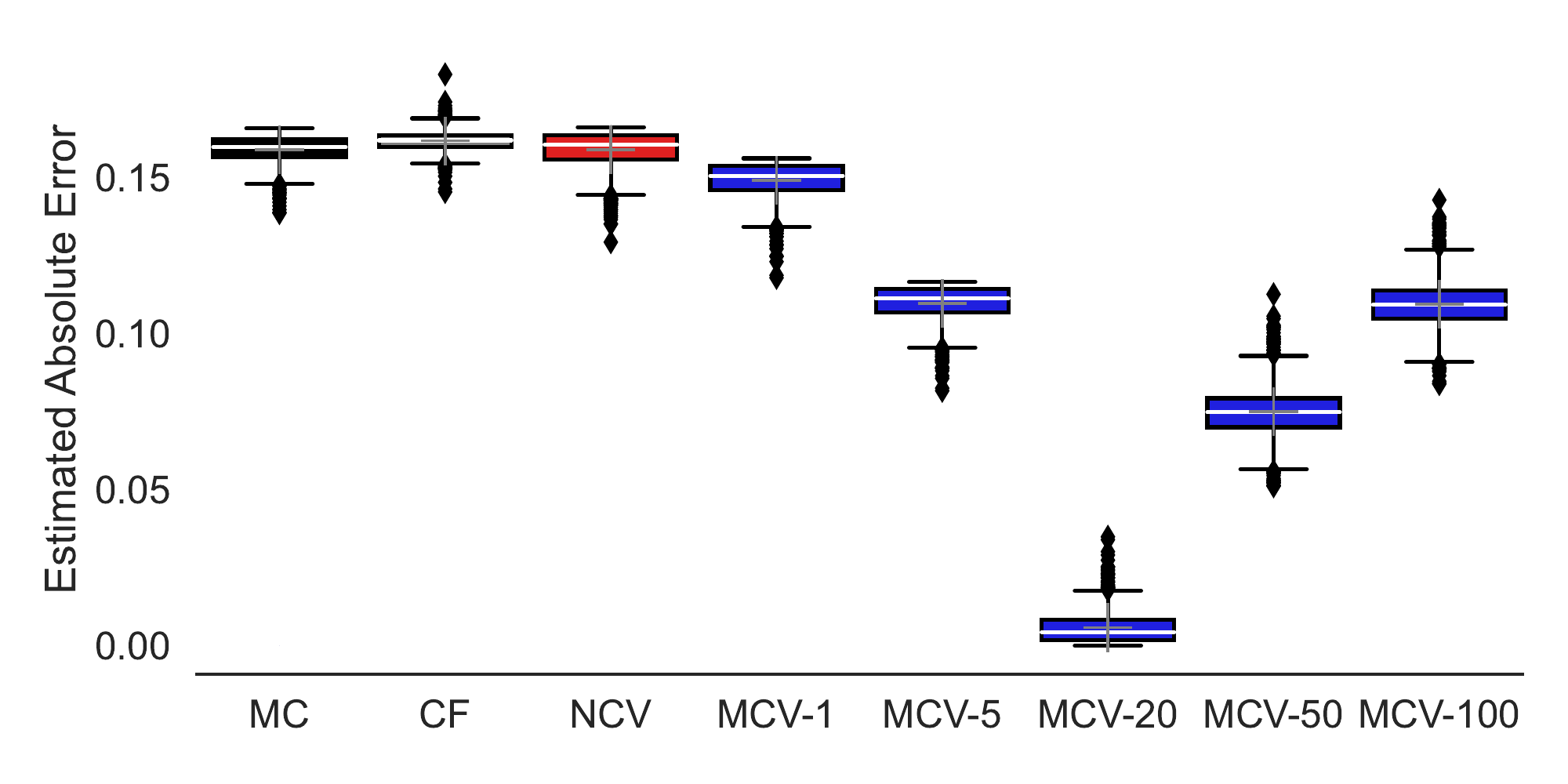}
    \caption{Effect of $L$: Estimated absolute errors over $\Ttest=1,000$ unseen states of the Sarcos anthropomorphic robot arm ($\text{\emph{CF}}$: Control functionals; $\text{\emph{NCV}}$: Neural-CVs; $\text{\emph{MCV-L}}$: Meta-CVs with $L$ inner steps).}
    \label{fig:sarcos_gammaprior_invest_innersteps}
\end{figure}

Our dataset is divided into two parts. The first part is used to obtain the posterior on Gaussian process hyperparameters (which is approximated through variational inference) and consists of $q=1,000$ data points. 
The second part includes $4,449$ data points, half of which are used to construct the Meta-CV and the other half is used to define a held-out test set of tasks for assessment. 
See \Cref{appdx:sarcos} for full experimental detail.

The results are presented in \Cref{fig:sarcos_gammaprior_invest_innersteps}.  We compare the performance of Meta-CVs with MC, CFs and Neural-CVs on $\Ttest=1,000$ unseen  tasks, where $N_t=4$ for each task. Although we do not have access to the exact value of these integrals, the value $y_t^*$ is an unbiased estimator, and this enables integration error to be unbiasedly estimated. We find that Meta-CVs are once again able to outperform competitors, but interestingly the performance improves significantly when the number of inner gradient steps $L>1$. Meanwhile, it is found empirically that there is a trade-off between remaining close to the Meta-CV, and specialising each CV to a specific task; see \citep{antoniou2018train} for a detailed discussion. In general, we would recommend to split the training set into a training set and a validation set, and choose the optimal value of $L$ (and other hyperparameters) on the validation set. This is known as a ``meta-validation'' process in the meta-learning literature.

\section{Theoretical Analysis}
\label{sec:theory}

The empirical results of the previous section demonstrate the advantage of leveraging the relationship between a large number of integration tasks. This section will focus on obtaining theoretical insight to guide the implementation of gradient-based optimisation within Meta-CVs.

Our analysis focuses on strategies for training of Meta-CVs. Recall that the (global) objective for learning a Meta-CV is 
\begin{talign}
\label{eq:obj_Fhat}
     \argmin_{\gamma \in \mathbb{R}^{p+1}} \; \; & \E_t \left[  \mathcal{J}_{t}(\gamma)   \right] , \\
     \mathcal{J}_{t}(\gamma) & :=  J_{Q_t} \left( \textsc{Update}_L \left(\gamma, \nabla_{\gamma} J_{S_t}\left(\gamma \right);\alpha \right)\right), \nonumber
\end{talign}
where in what follows $\textsc{Update}_L$ is gradient descent with $L$ steps and inner-step size $\alpha$. 
To proceed, we make the following assumptions:
\begin{assumption}
\label{assum:cv_new1}
    For each $t$ and $x\in D_t$, $\gamma \mapsto g(x;\gamma)$ and $\gamma \mapsto \nabla_\gamma g(x;\gamma)$ are bounded and Lipschitz.
\end{assumption}

\begin{assumption}
\label{assum:cv_new2}
    For each $t$ and $x\in D_t$, $\gamma \mapsto \nabla_\gamma g(x;\gamma) \nabla_\gamma g(x;\gamma)^\top - \nabla_\gamma^2 g(x;\gamma)$ is bounded and Lipschitz.
\end{assumption}

\Cref{assum:cv_new1} can in principle be satisfied by the Stein-based CVs introduced in \Cref{sec:background}, since it concerns the behaviour of $g(x;\gamma)$ and $\nabla_\gamma g(x;\gamma)$ as $\gamma$, rather than $x$, is varied (recall that, as a function of $x$, Stein-based CVs are usually unbounded). For \Cref{assum:cv_new2}, we note that $\nabla_\gamma g(x;\gamma) \nabla_\gamma g(x;\gamma)^\top$ is a popular low-rank approximation to the Hessian $\nabla_\gamma^2 g(x;\gamma)$, so \Cref{assum:cv_new2} explicitly requires this low-rank approximation to be reasonably good.

The following theorem, which builds on the work of \citet{ji2022theoretical_ms_gbml}, establishes conditions under which Algorithm~\ref{alg:meta_neural_cv_train} can find an $\epsilon$-first order stationary point of the meta-learning objective function \Cref{eq:obj_Fhat}, for any $\epsilon > 0$.

\begin{theorem}
\label{theoCV:cvepsionexits}
 Let $\hat{\gamma}_{\text{\normalfont meta}}$ be the output of Algorithm~\ref{alg:meta_neural_cv_train} with gradient descent steps, using the meta-step-sizes $\eta_1,\ldots,\eta_{\nepoch}$, the inner-step size $\alpha$ and batch size $B$ proposed in Theorem 9 and Corollary 10 of \citet{ji2022theoretical_ms_gbml}. Then, under Assumptions \ref{assum:cv_new1}, \ref{assum:cv_new2}:
\begin{talign*}
    \mathbb{E} [\| \E_t [\nabla \mathcal{J}_t(\hat{\gamma}_{\text{\normalfont meta}})] \|_2] = \mathcal{O} \left( \sqrt{ \frac{1}{\nepoch} + \frac{1}{B} } ~\right) ,
\end{talign*}
where the outer expectation is with respect to sampling of the mini-batches of tasks in Algorithm~\ref{alg:meta_neural_cv_train}. 
\end{theorem}
The proof is contained in Appendix \ref{appdx:proof_thereom1}.
If we take $B \geq C_B \epsilon^{-2}$ with $C_B$ a large constant, the theorem shows that, with at most $\nepoch =\O(\nicefrac{1}{\epsilon^2})$ meta iterations, the output $\hat{\gamma}_{\epsilon}$ of Algorithm~\ref{alg:meta_neural_cv_train} satisfies $\mathbb{E}[ \| \E_t[\nabla \mathcal{J}_t(\hat{\gamma}_{\epsilon})] \|] = \O(\epsilon)$. The requirements on the step-sizes and batch size are inherited from  \cite{ji2022theoretical_ms_gbml}, are spelled out in \Cref{appdx:previous_results}, and provide guiding insight into the practical side of training of Meta-CVs, e.g. theoretically optimal meta-step-sizes $\eta_1, \ldots, \eta_{\nepoch}$ and inner-step size $\alpha$.
For Neural CVs it is difficult to go beyond \Cref{theoCV:cvepsionexits}, since for one thing there will not be a unique $\gamma_{\text{meta}}$ in general.
However, for simpler CVs, such as those based on polynomial regression \citep{assaraf1999_zerovariance_principleforMCI,mira2013_zerovariance_MCMC,Papamarkou2014,Friel2014,South2019}, it is reasonable to assume a unique $\gamma_{\text{meta}}$ and convexity of the Meta-CV objective around this point.
In these scenarios, the following corollary shows that $\hat{\gamma}_{\epsilon}$ is typically close to the minimiser of the task-specific objective functional.

\begin{corollary}
\label{theoCV:from_r_to_ri}
    Under the setting of \Cref{theoCV:cvepsionexits}, further suppose that there exists $\mu > 0$ such that for all $t$ and all $\gamma$,  $\nabla^2 J_{Q_t}(\gamma) \succeq \mu I_{p+1}$ where $I_{p+1}$ is an identity matrix of size $p+1$.
    Then there exist constants $C_1 , C_2>0$ such that
    \begin{talign*}
         \E[\E_{t} [\|\hat{\gamma}_{\epsilon} - \gamma^*_t\|_2] ]\leq \frac{C_{1}}{\mu}\epsilon + \frac{C_2}{\mu},
    \end{talign*}
    where $\gamma^*_t $ is the (unique) minimiser of $\gamma \mapsto J_{Q_t}(\gamma)$, and here again the outer expectation is with respect to sampling of the mini-batches of tasks in Algorithm~\ref{alg:meta_neural_cv_train}.
\end{corollary}
The proof is contained in Appendix \ref{appdx:proof_thereom2}. These results justify the use of Algorithm 1 to train the Meta-CV and task-specific CVs. In particular, they provide insight into step size selection, and establish explicit conditions on the form of CV $g(x;\gamma)$ that can be successfully trained using the methodology that we have proposed.

\section{Conclusion}

This paper introduced Meta-CVs, an extension of existing CV methods that brings meta-learning to bear on MC and MCMC. 
More precisely, our method can achieve significant variance reduction when the number of samples per integration task is small, but a large number $T$ of similar tasks are available. In addition, most of the computational cost is an offline cost for identifying a Meta-CV, and CVs for new integration tasks can be identified with minimal additional computational cost.

Although our algorithm is scalable in $T$ and $N_t$, the computational cost for training the Meta-CV can still be significant when dealing with flexible CVs, such as Neural CVs. 
For example, computational complexity scales as $\O(p^2)$ in the number of parameters $p$ in the CV. This prevents us from using very large neural networks,  which could limit performance on more challenging integration tasks. 
First-order or Hessian-free meta-learning algorithms \citep{fallah2020convergence_gbml} are therefore a promising direction for future work.

Alternatively, online meta-learning algorithms \citep{finn2019online_maml} could be adapted to CVs. These could be particularly powerful for cases where integration tasks arrive sequentially and the Meta-CV cannot be computed offline. Examples includes application areas where sequential importance sampling and sequential MC-type algorithms \citep{doucet2000sequential, doucet2001sequential} are currently being used, such as in the context of state-space models. 

Finally, it is also possible to further extend our theoretical analysis of Meta-CVs. The current convergence rate in $\nepoch$ of Meta-CVs aligns with \citep{fallah2020convergence_gbml, ji2022theoretical_ms_gbml}. Future work could extend the theoretical analysis of Meta-CVs from a information-theoretic aspect \citep{chen2021generalization} or towards a faster rate \citep{riou2023bayes} with additional conditions.



\begin{acknowledgements}
    The authors would like to thank Kaiyu Li for sharing some of her code for the boundary value ODE example.
    ZS was supported under the EPSRC grant [EP/R513143/1] and The Alan Turing Institute’s Enrichment Scheme. CJO and FXB were supported by the Lloyd’s Register Foundation Programme on Data-Centric Engineering and The Alan Turing Institute under the EPSRC grant [EP/N510129/1]. 
    CJO was supported by the EPSRC grant [EP/W019590/1].
\end{acknowledgements}

\bibliography{sun_447}

\begin{thebibliography}{69}
\providecommand{\natexlab}[1]{#1}
\providecommand{\url}[1]{\texttt{#1}}
\expandafter\ifx\csname urlstyle\endcsname\relax
  \providecommand{\doi}[1]{doi: #1}\else
  \providecommand{\doi}{doi: \begingroup \urlstyle{rm}\Url}\fi

\bibitem[Alexopoulos et~al.(2023)Alexopoulos, Dellaportas, and
  Titsias]{Alexopoulos2023}
A.~Alexopoulos, P.~Dellaportas, and M.~K. Titsias.
\newblock {Variance reduction for Metropolis–Hastings samplers}.
\newblock \emph{Stat. Comput.}, 33\penalty0 (6), 2023.

\bibitem[Anastasiou et~al.(2023)Anastasiou, Barp, Briol, Ebner, Ghaderinezhad,
  Gorham, Gretton, Ley, Liu, et~al.]{Anastasiou2021}
A.~Anastasiou, A.~Barp, F-X. Briol, R.~E. Ebner, B.and~Gaunt, F.~Ghaderinezhad,
  J.~Gorham, A.~Gretton, C.~Ley, Q.~Liu, et~al.
\newblock Stein’s method meets computational statistics: a review of some
  recent developments.
\newblock \emph{Stat. Sci.}, 38\penalty0 (1):\penalty0 120--139, 2023.

\bibitem[Andrychowicz et~al.(2016)Andrychowicz, Denil, Gomez, Hoffman, Pfau,
  Schaul, Shillingford, and De~Freitas]{andrychowicz2016_l2l_GD_by_GD}
M.~Andrychowicz, M.~Denil, S.~Gomez, M.~W. Hoffman, D.~Pfau, T.~Schaul,
  B.~Shillingford, and N.~De~Freitas.
\newblock Learning to learn by gradient descent by gradient descent.
\newblock \emph{NeurIPS}, 2016.

\bibitem[Antoniou et~al.(2019)Antoniou, Edwards, and
  Storkey]{antoniou2018train}
A.~Antoniou, H.~Edwards, and A.~Storkey.
\newblock How to train your maml.
\newblock In \emph{ICLR}, 2019.

\bibitem[Assaraf and Caffarel(1999)]{assaraf1999_zerovariance_principleforMCI}
R.~Assaraf and M.~Caffarel.
\newblock Zero-variance principle for {M}onte {C}arlo algorithms.
\newblock \emph{Phys. Rev. Lett.}, 83\penalty0 (23):\penalty0 4682, 1999.

\bibitem[Baker et~al.(2019)Baker, Fearnhead, Fox, and Nemeth]{Baker2018}
J.~Baker, P.~Fearnhead, E.~B. Fox, and C.~Nemeth.
\newblock {Control variates for stochastic gradient MCMC}.
\newblock \emph{Stat. Comput.}, 29:\penalty0 599--615, 2019.

\bibitem[Barp et~al.(2022)Barp, Oates, Porcu, and Girolami]{Barp2018}
A.~Barp, C.~J. Oates, E.~Porcu, and M.~Girolami.
\newblock {A Riemannian–Stein kernel method}.
\newblock \emph{Bernoulli}, 28\penalty0 (4):\penalty0 2181--2208, 2022.

\bibitem[Belomestny et~al.(2020)Belomestny, Iosipoi, Moulines, Naumov, and
  Samsonov]{Belomestny2019}
D.~Belomestny, L.~Iosipoi, E.~Moulines, A.~Naumov, and S.~Samsonov.
\newblock {Variance reduction for Markov chains with application to MCMC}.
\newblock \emph{Stat. Comput.}, 30:\penalty0 973--997, 2020.

\bibitem[Belomestny et~al.(2021)Belomestny, Iosipoi, Moulines, Naumov, and
  Samsonov]{Belomestny2021}
D.~Belomestny, L.~Iosipoi, E.~Moulines, Al. Naumov, and S.~Samsonov.
\newblock {Variance reduction for dependent sequences with applications to
  stochastic gradient MCMC}.
\newblock \emph{SIAM-ASA J. Uncertain.}, 9\penalty0 (1):\penalty0 507--535,
  2021.

\bibitem[Bolker(2008)]{bolker2008ecological}
B.~M. Bolker.
\newblock {Ecological models and data in R}.
\newblock In \emph{Ecological Models and Data in R}. Princeton University
  Press, 2008.

\bibitem[Boyd et~al.(2004)Boyd, Boyd, and Vandenberghe]{boyd2004convex}
S.~Boyd, S.~P. Boyd, and L.~Vandenberghe.
\newblock \emph{Convex optimization}.
\newblock Cambridge university press, 2004.

\bibitem[Brauer(2017)]{brauer2017mathematical}
F.~Brauer.
\newblock Mathematical epidemiology: Past, present, and future.
\newblock \emph{Infect. Dis. Model.}, 2\penalty0 (2):\penalty0 113--127, 2017.

\bibitem[Carpenter et~al.(2017)Carpenter, Gelman, Hoffman, Lee, Goodrich,
  Betancourt, Brubaker, Guo, Li, and Riddell]{carpenter2017stan}
B.~Carpenter, A.~Gelman, M.~D. Hoffman, D.~Lee, B.~Goodrich, M.~Betancourt,
  M.~Brubaker, J.~Guo, P.~Li, and A.~Riddell.
\newblock Stan: A probabilistic programming language.
\newblock \emph{J. Stat. Softw.}, 76\penalty0 (1), 2017.

\bibitem[Chen et~al.(2021)Chen, Shui, and Marchand]{chen2021generalization}
Q.~Chen, C.~Shui, and M.~Marchand.
\newblock Generalization bounds for meta-learning: An information-theoretic
  analysis.
\newblock \emph{NeurIPS}, 34:\penalty0 25878--25890, 2021.

\bibitem[Dellaportas and Kontoyiannis(2012)]{Dellaportas2012}
P.~Dellaportas and I~Kontoyiannis.
\newblock {Control variates for estimation based on reversible Markov chain
  Monte Carlo samplers}.
\newblock \emph{J. R. Stat. Soc. Series B}, 74\penalty0 (1):\penalty0 133--161,
  2012.

\bibitem[Demange-Chryst et~al.(2022)Demange-Chryst, Bachoc, and
  Morio]{Demange-Chryst2022}
J.~Demange-Chryst, F.~Bachoc, and J.~Morio.
\newblock {Efficient estimation of multiple expectations with the same sample
  by adaptive importance sampling and control variates}.
\newblock \emph{arXiv:2212.00568}, 2022.

\bibitem[Doucet et~al.(2000)Doucet, Godsill, and Andrieu]{doucet2000sequential}
A.~Doucet, S.~Godsill, and C.~Andrieu.
\newblock On sequential monte carlo sampling methods for bayesian filtering.
\newblock \emph{Stat. Comput.}, 10:\penalty0 197--208, 2000.

\bibitem[Doucet et~al.(2001)Doucet, De~Freitas, and
  Gordon]{doucet2001sequential}
A.~Doucet, N.~De~Freitas, and N.~J. Gordon.
\newblock \emph{Sequential Monte Carlo methods in practice}, volume~1.
\newblock Springer, 2001.

\bibitem[Fallah et~al.(2020)Fallah, Mokhtari, and
  Ozdaglar]{fallah2020convergence_gbml}
A.~Fallah, A.~Mokhtari, and A.~Ozdaglar.
\newblock On the convergence theory of gradient-based model-agnostic
  meta-learning algorithms.
\newblock In \emph{AISTATS}. PMLR, 2020.

\bibitem[Finn et~al.(2017)Finn, Abbeel, and Levine]{icml_2017_finn17a_MAML}
C.~Finn, P.~Abbeel, and S.~Levine.
\newblock Model-agnostic meta-learning for fast adaptation of deep networks.
\newblock In \emph{ICML}, 2017.

\bibitem[Finn et~al.(2018)Finn, Xu, and
  Levine]{nips_2018_fin_probabilisticMAML}
C.~Finn, K.~Xu, and S.~Levine.
\newblock Probabilistic model-agnostic meta-learning.
\newblock In \emph{NeurIPS}, 2018.

\bibitem[Finn et~al.(2019)Finn, Rajeswaran, Kakade, and
  Levine]{finn2019online_maml}
C.~Finn, A.~Rajeswaran, S.~Kakade, and S.~Levine.
\newblock Online meta-learning.
\newblock In \emph{ICML}, 2019.

\bibitem[Friel et~al.(2014)Friel, Mira, and Oates]{Friel2014}
N.~Friel, A.~Mira, and C.~J. Oates.
\newblock {Exploiting multi-core architectures for reduced-variance estimation
  with intractable likelihoods}.
\newblock \emph{Bayesian Anal.}, 11\penalty0 (1):\penalty0 215--245, 2014.

\bibitem[Gessner et~al.(2019)Gessner, Gonzalez, and Mahsereci]{Gessner2019}
A.~Gessner, J.~Gonzalez, and M.~Mahsereci.
\newblock {Active multi-information source Bayesian quadrature}.
\newblock In \emph{UAI}, 2019.

\bibitem[Giles(2015)]{giles2015multilevel}
M.~Giles.
\newblock {Multilevel Monte Carlo methods}.
\newblock \emph{Acta Numer.}, 24:\penalty0 259--328, 2015.

\bibitem[Glynn and Igelhart(1989)]{Glynn1989}
P.~Glynn and D.~Igelhart.
\newblock {Importance sampling for stochastic simulations}.
\newblock \emph{Management Science}, 35\penalty0 (1367-1392), 1989.

\bibitem[Grant et~al.(2018)Grant, Finn, Levine, Darrell, and
  Griffiths]{iclr_2018_grant_recasting_HBMAML}
E.~Grant, C.~Finn, S.~Levine, T.~Darrell, and T.~Griffiths.
\newblock Recasting gradient-based meta-learning as hierarchical {Bayes}.
\newblock In \emph{ICML}, 2018.

\bibitem[Grathwohl et~al.(2018)Grathwohl, Choi, Wu, Roeder, and
  Duvenaud]{Grathwohl2017}
W.~Grathwohl, D.~Choi, Y.~Wu, G.~Roeder, and D.~Duvenaud.
\newblock {Backpropagation through the void: Optimizing control variates for
  black-box gradient estimation}.
\newblock In \emph{ICLR}, 2018.

\bibitem[Green et~al.(2015)Green, Latuszyski, Pereyra, and Robert]{Green2015}
P.~Green, K.~Latuszyski, M.~Pereyra, and C.~Robert.
\newblock {Bayesian computation: a summary of the current state, and samples
  backwards and forwards}.
\newblock \emph{Stat. Comput.}, 25:\penalty0 835--862, 2015.

\bibitem[Grefenstette et~al.(2019)Grefenstette, Amos, Yarats, Htut, Molchanov,
  Meier, Kiela, Cho, and Chintala]{grefenstette2019generalized_maml}
E.~Grefenstette, B.~Amos, D.~Yarats, P.~Htut, A.~Molchanov, F.~Meier, D.~Kiela,
  K.~Cho, and S.~Chintala.
\newblock Generalized inner loop meta-learning.
\newblock \emph{arXiv:1910.01727}, 2019.

\bibitem[Hewitt(1921)]{hewitt1921conservation}
C.~Hewitt.
\newblock \emph{The conservation of the wild life of Canada}.
\newblock New York: C. Scribner, 1921.

\bibitem[Hickernell et~al.(2005)Hickernell, Lemieux, and Owen]{Hickernell2005}
F.J. Hickernell, C.~Lemieux, and A.~B. Owen.
\newblock {Control variates for quasi-Monte Carlo}.
\newblock \emph{Stat. Sci.}, 20\penalty0 (1):\penalty0 1--31, 2005.

\bibitem[Ji et~al.(2022)Ji, Yang, and Liang]{ji2022theoretical_ms_gbml}
K.~Ji, J.~Yang, and Y.~Liang.
\newblock Theoretical convergence of multi-step model-agnostic meta-learning.
\newblock \emph{J. Mach. Learn. Res.}, 23:\penalty0 29--1, 2022.

\bibitem[Kingma and Ba(2015)]{Kingma2015}
D.~P. Kingma and J.~L. Ba.
\newblock {Adam: A method for stochastic optimization}.
\newblock In \emph{ICLR}, 2015.

\bibitem[Kingma and Welling(2014)]{kingma2013_VAE}
D.~P. Kingma and M.~Welling.
\newblock Auto-encoding variational bayes.
\newblock In \emph{ICLR}, 2014.

\bibitem[Krumscheid and Nobile(2018)]{krumscheid2018multilevel}
S.~Krumscheid and F.~Nobile.
\newblock Multilevel monte carlo approximation of functions.
\newblock \emph{SIAM-ASA J. Uncertain.}, 6\penalty0 (3):\penalty0 1256--1293,
  2018.

\bibitem[Kucukelbir et~al.(2017)Kucukelbir, Tran, Ranganath, Gelman, and
  Blei]{kucukelbir2017automatic_VL}
A.~Kucukelbir, D.~Tran, R.~Ranganath, A.~Gelman, and D.~M. Blei.
\newblock Automatic differentiation variational inference.
\newblock \emph{J. Mach. Learn. Res.}, 2017.

\bibitem[Lalchand and Rasmussen(2020)]{lalchand2020approximate_FB_GP}
V.~Lalchand and C.~E. Rasmussen.
\newblock {Approximate inference for fully Bayesian Gaussian process
  regression}.
\newblock In \emph{AABI}, pages 1--12. PMLR, 2020.

\bibitem[Leluc et~al.(2021)Leluc, Portier, and
  Segers]{leluc2021_cv_selection_for_montecarlo_integration}
R.~Leluc, F.~Portier, and J.~Segers.
\newblock Control variate selection for monte carlo integration.
\newblock \emph{Stat. Comput.}, 31\penalty0 (4):\penalty0 1--27, 2021.

\bibitem[Li et~al.(2023)Li, Giles, Karvonen, Guillas, and Briol]{Li2022}
K.~Li, D.~Giles, T.~Karvonen, S.~Guillas, and F-X. Briol.
\newblock {Multilevel Bayesian quadrature}.
\newblock In \emph{AISTATS}, pages 1845--1868, 2023.

\bibitem[Liu et~al.(2018)Liu, Feng, Mao, Zhou, Peng, and Liu]{Liu2018}
H.~Liu, Y.~Feng, Y.~Mao, D.~Zhou, J.~Peng, and Q.~Liu.
\newblock {Action-dependent control variates for policy optimization via
  stein's identity}.
\newblock In \emph{ICLR}, 2018.

\bibitem[Liu et~al.(2019)Liu, Socher, and Xiong]{liu2019taming}
H.~Liu, R.~Socher, and C.~Xiong.
\newblock Taming maml: Efficient unbiased meta-reinforcement learning.
\newblock In \emph{ICML}. PMLR, 2019.

\bibitem[Longstaff and Schwartz(2001)]{Longstaff2001}
F.~A. Longstaff and E.~S. Schwartz.
\newblock {Valuing american options by simulation: A simple least-squares
  approach}.
\newblock \emph{Rev. Financ. Stud.}, 14\penalty0 (1):\penalty0 113--147, 2001.

\bibitem[Lotka(1927)]{lotka1927fluctuations}
A.~Lotka.
\newblock Fluctuations in the abundance of a species considered mathematically.
\newblock \emph{Nature}, 119\penalty0 (2983):\penalty0 12--12, 1927.

\bibitem[Madras and Piccioni(1999)]{Madras1999}
N.~Madras and M.~Piccioni.
\newblock {Importance sampling for families of distributions}.
\newblock \emph{Ann. Appl. Probab.}, 9\penalty0 (4):\penalty0 1202--1225, 1999.

\bibitem[Mira et~al.(2013)Mira, Solgi, and
  Imparato]{mira2013_zerovariance_MCMC}
A.~Mira, R.~Solgi, and D.~Imparato.
\newblock {Zero variance Markov chain Monte Carlo for Bayesian estimators}.
\newblock \emph{Stat. Comput.}, 23\penalty0 (5):\penalty0 653--662, 2013.

\bibitem[M{\"{u}}ller et~al.(2020)M{\"{u}}ller, Rousselle, Nov{\'{a}}k, and
  Keller]{Muller2020}
T.~M{\"{u}}ller, F.~Rousselle, J.~Nov{\'{a}}k, and A.~Keller.
\newblock {Neural control variates}.
\newblock \emph{ACM Trans. Graph.}, 39\penalty0 (6):\penalty0 1--19, 2020.

\bibitem[Oates and Girolami(2016)]{Oates2016CFQMC}
C.~J Oates and M.~Girolami.
\newblock {Control functionals for quasi-Monte Carlo integration}.
\newblock In \emph{AISTATS}, 2016.

\bibitem[Oates et~al.(2016)Oates, Papamarkou, and Girolami]{Oates2016thermo}
C.~J. Oates, T.~Papamarkou, and M.~Girolami.
\newblock {The controlled thermodynamic integral for Bayesian model
  comparison}.
\newblock \emph{J. Am. Stat. Assoc.}, 111\penalty0 (514):\penalty0 634--645,
  2016.

\bibitem[Oates et~al.(2017)Oates, Girolami, and
  Chopin]{oates2017_CF_for_MonteCarloIntegration}
C.~J. Oates, M.~Girolami, and N.~Chopin.
\newblock {Control functionals for Monte Carlo integration}.
\newblock \emph{J. R. Stat. Soc. Series B}, 79\penalty0 (3):\penalty0 695--718,
  2017.

\bibitem[Oates et~al.(2019)Oates, Cockayne, Briol, and
  Girolami]{oates2019convergence}
C.~J. Oates, J.~Cockayne, F-X. Briol, and M.~Girolami.
\newblock Convergence rates for a class of estimators based on {S}tein’s
  method.
\newblock \emph{Bernoulli}, 25\penalty0 (2):\penalty0 1141--1159, 2019.

\bibitem[Ott et~al.(2023)Ott, Tiemann, Hennig, and Briol]{Ott2023}
K.~Ott, M.~Tiemann, P.~Hennig, and F-X. Briol.
\newblock {Bayesian numerical integration with neural networks}.
\newblock \emph{arXiv:2305.13248}, 2023.

\bibitem[Paisley et~al.(2012)Paisley, Blei, and Jordan]{paisley2012variational}
J.~Paisley, D.~M. Blei, and M.~I. Jordan.
\newblock Variational bayesian inference with stochastic search.
\newblock In \emph{ICML}, 2012.

\bibitem[Papamarkou et~al.(2014)Papamarkou, Mira, and Girolami]{Papamarkou2014}
T.~Papamarkou, A.~Mira, and M.~Girolami.
\newblock {Zero variance differential geometric Markov chain Monte Carlo
  algorithms}.
\newblock \emph{Bayesian Anal.}, 9\penalty0 (1):\penalty0 97--128, 2014.

\bibitem[Paszke et~al.(2019)Paszke, Gross, Massa, Lerer, Bradbury, Chanan,
  Killeen, Lin, Gimelshein, Antiga, et~al.]{paszke2019pytorch}
A.~Paszke, S.~Gross, F.~Massa, A.~Lerer, J.~Bradbury, G.~Chanan, T.~Killeen,
  Z.~Lin, N.~Gimelshein, L.~Antiga, et~al.
\newblock Pytorch: An imperative style, high-performance deep learning library.
\newblock \emph{NeurIPS}, 32, 2019.

\bibitem[Peherstorfer et~al.(2018)Peherstorfer, Willcox, and
  Gunzburger]{Peherstorfer2018}
B.~Peherstorfer, K.~Willcox, and M.~Gunzburger.
\newblock {Survey of multifidelity methods in uncertainty propagation,
  inference, and optimization}.
\newblock \emph{SIAM Review}, 60\penalty0 (3):\penalty0 550--591, 2018.

\bibitem[Rasmussen and Williams(2006)]{rasmussen2003_GPML}
C.~E. Rasmussen and C.~K.~I. Williams.
\newblock \emph{Gaussian processes for machine learning}, volume~1.
\newblock Springer, 2006.

\bibitem[Riou et~al.(2023)Riou, Alquier, and
  Ch{\'e}rief-Abdellatif]{riou2023bayes}
C.~Riou, P.~Alquier, and B-E. Ch{\'e}rief-Abdellatif.
\newblock Bayes meets bernstein at the meta level: an analysis of fast rates in
  meta-learning with pac-bayes.
\newblock \emph{arXiv preprint arXiv:2302.11709}, 2023.

\bibitem[Shi et~al.(2022)Shi, Zhou, Hwang, Titsias, and Mackey]{Shi2022}
J.~Shi, Y.~Zhou, J.~Hwang, M.~K. Titsias, and L.~Mackey.
\newblock {Gradient estimation with discrete Stein operators}.
\newblock In \emph{NeurIPS}, 2022.

\bibitem[Si et~al.(2021)Si, Oates, Duncan, Carin, and Briol]{Si2020}
S.~Si, C.~J. Oates, A.~B. Duncan, L.~Carin, and F-X. Briol.
\newblock {Scalable control variates for Monte Carlo methods via stochastic
  optimization}.
\newblock \emph{Proceedings of the 14th Conference on Monte Carlo and
  Quasi-Monte Carlo Methods. arXiv:2006.07487}, 2021.

\bibitem[South et~al.(2022{\natexlab{a}})South, Karvonen, Nemeth, Girolami, and
  Oates]{south2020_Semi_Exact_CF}
L.~F. South, T.~Karvonen, C.~Nemeth, M.~Girolami, and C.~J. Oates.
\newblock {Semi-exact control functionals from Sard's method}.
\newblock \emph{Biometrika}, 2022{\natexlab{a}}.

\bibitem[South et~al.(2022{\natexlab{b}})South, Oates, Mira, and
  Drovandi]{South2019}
L.~F. South, C.~J. Oates, A.~Mira, and C.~Drovandi.
\newblock Regularized zero-variance control variates.
\newblock \emph{Bayesian Anal.}, 1\penalty0 (1):\penalty0 1--24,
  2022{\natexlab{b}}.

\bibitem[South et~al.(2022{\natexlab{c}})South, Riabiz, Teymur, and
  Oates]{South2022}
L.~F. South, M.~Riabiz, O.~Teymur, and C.~J. Oates.
\newblock {Post-Processing of MCMC}.
\newblock \emph{Annu. Rev. Stat. Appl.}, 2022{\natexlab{c}}.

\bibitem[Sun et~al.(2021{\natexlab{a}})Sun, Barp, and Briol]{sun2021_vvCV}
Z.~Sun, A.~Barp, and F-X. Briol.
\newblock {Vector-Valued Control Variates}.
\newblock \emph{arXiv:2109.08944, to appear at ICML 2023}, 2021{\natexlab{a}}.

\bibitem[Sun et~al.(2021{\natexlab{b}})Sun, Wu, Li, Yang, and
  Xue]{sun2021amortized}
Z.~Sun, J.~Wu, X.~Li, W.~Yang, and J-H. Xue.
\newblock {Amortized Bayesian Prototype Meta-learning: A new probabilistic
  meta-learning approach to few-shot image classification}.
\newblock In \emph{AISTATS}, pages 1414--1422. PMLR, 2021{\natexlab{b}}.

\bibitem[Wan et~al.(2019)Wan, Zhong, Xiong, and Zhu]{zhu2018neural_CV}
R.~Wan, M.~Zhong, H.~Xiong, and Z.~Zhu.
\newblock Neural control variates for variance reduction.
\newblock \emph{ECML PKDD}, page 533–547, 2019.

\bibitem[Wang et~al.(2013)Wang, Chen, Smola, and Xing]{wang2013variance}
C.~Wang, X.~Chen, A.~J. Smola, and E.~P. Xing.
\newblock Variance reduction for stochastic gradient optimization.
\newblock \emph{NeurIPS}, 2013.

\bibitem[Xi et~al.(2018)Xi, Briol, and
  Girolami]{ICML2018_BQforMultipleRelatedIntegrals}
X.~Xi, F-X. Briol, and M.~Girolami.
\newblock {B}ayesian quadrature for multiple related integrals.
\newblock In \emph{ICML}, 2018.

\bibitem[Yoon et~al.(2018)Yoon, Kim, Dia, Kim, Bengio, and
  Ahn]{NIPS_2018_yoon_BayesianMAML}
J.~Yoon, T.~Kim, O.~Dia, S.~Kim, Y.~Bengio, and S.~Ahn.
\newblock Bayesian model-agnostic meta-learning.
\newblock In \emph{NeurIPS}, 2018.

\end{thebibliography}

\newpage
\onecolumn
\appendix

{
\begin{center}
\LARGE
    \vspace{5mm}
    \textbf{Appendix}
    \vspace{5mm}
\end{center}
}

In \Cref{appdx:proofs}, we provide the proof of the theoretical results stated in the main text. In \Cref{appdx:experimental_details}, we provide more details on the implementation of Neural-CVs and Meta-CVs, together with the full experimental protocol.

\section{Proof of Theorems}
\label{appdx:proofs}

In this section, we will firstly review the assumptions and theorems in \citep{ji2022theoretical_ms_gbml} in \Cref{appdx:previous_results} as the proof of the theorems follows the results of \citep{ji2022theoretical_ms_gbml}. We then give the proof of \Cref{theoCV:cvepsionexits} in \Cref{appdx:proof_thereom1} and proof of \Cref{theoCV:from_r_to_ri} in \Cref{appdx:proof_thereom2}.

\subsection{Convergence of Model-Agnostic Meta-Learning}
\label{appdx:previous_results}

\citet{ji2022theoretical_ms_gbml} analysed the convergence of model-agnostic meta-learning, as we will adapt their results to the training of CVs. 
Letting $O_t$ be either $S_t$ or $Q_t$, and phrasing in terms of the notation and setting used in this work, the assumptions of \citep{ji2022theoretical_ms_gbml} are:
\begin{enumerate}
\item[(A1)] $\min_t \inf_{\gamma} J_{O_t}(\gamma) >-\infty ;$
\item[(A2)] $ \chi := \max_t \sup_{\gamma \neq \zeta} \frac{ \|\nabla_\gamma J_{O_t}(\gamma) - \nabla_\zeta J_{O_t}(\zeta) \|_2 }{ \|\gamma - \zeta \|_2 } < \infty ;$
\item[(A3)] $ \rho := \max_t \sup_{\gamma \neq \zeta} \frac{ \|\nabla_\gamma^2 J_{O_t}(\gamma) - \nabla_\zeta^2 J_{O_t}(\zeta) \|_2 }{ \|\gamma - \zeta \|_2 } < \infty ;$
\item[(A4)] $ \sigma^2 := \max_t \sup_\gamma \| \nabla_\gamma J_{O_t}(\gamma) \|^2_2 < \infty ;$
\item[(A5)] $ b_t := \sup_\gamma \|J_{S_t}(\gamma) - J_{Q_t}(\gamma)\|_2 < \infty .$
\end{enumerate}

\begin{theorem}[Theorem 9 and Corollary 10 \citep{ji2022theoretical_ms_gbml}]
\label{theo:theo9_coro10_in_conver_gbml}
 Let the above assumptions (A1) to (A5) hold. Then, with a meta step-size $\eta_i =\frac{1}{80\chi_{\eta_i}}$ for $i = 1, \ldots, \nepoch$ and $\alpha = \frac{1}{8 \chi L}$ in Algorithm \ref{alg:meta_neural_cv_train} , we attain a solution $\hat{\gamma}_{\text{\normalfont meta}}$ such that
 \begin{talign*}
     \E \|\E_t [ \nabla \J_t( \hat{\gamma}_{\text{\normalfont meta}} ) ] \|_2 = \O\left( \frac{1}{\nepoch} + \frac{\sigma^2}{B} + \sqrt{\frac{1}{\nepoch} + \frac{\sigma^2}{B}} \right),
 \end{talign*}
 where $\chi_{\eta_i} = (1+\alpha \chi)^{2L} + C_b b + C_\chi \E_t[\|\nabla J_{Q_t}( \hat{\gamma}_{\text{\normalfont meta}} )\|_2]$,
with $b = \mathbb{E}_{t}[b_t]$ and
$C_b =C_\chi =(\alpha \rho+ \nicefrac{\rho}{\chi}(1+\alpha \chi)^{L-1})(1+\alpha \chi)^{2L}$.
\end{theorem}

\begin{lemma}[Lemma 19 \citep{ji2022theoretical_ms_gbml}]
\label{lemma:lemma19_in_conver_gbml}
Under assumptions (A1) - (A5), for any $t$ and any $\gamma \in \R^{p+1}$, we have 
\begin{talign*}
    \| \E_{t} [ \nabla J_{Q_{t}}(\gamma) ] \|_2 \leq\frac{1}{C_1'}\|\E_{t} [ \nabla \J_t(\gamma) ] \|_2 + \frac{C_2'}{C_1'},
\end{talign*}
where $C_1'> 0$ and $C_2' > 0$ are constants given $C_1' = 2-(1+\alpha \chi)^{2L}$ and $C_2' = ((1+\alpha \chi)^{2L} - 1)\sigma + (1+\alpha \chi)^{L}((1+\alpha \chi)^L-1)b$.
\end{lemma}

\subsection{Proof of Theorem~\ref{theoCV:cvepsionexits}}
\label{appdx:proof_thereom1}
To prove \Cref{theoCV:cvepsionexits}, we firstly derive three useful propositions (P1-P3) based on our \Cref{assum:cv_new1} and \Cref{assum:cv_new2} in \Cref{sec:theory}, and then give the proof based on the above results from \citep{ji2022theoretical_ms_gbml}.

For each task $t$, we claim that
\begin{enumerate}
\item[(P1)] $ \sup_{\gamma \neq \zeta} \frac{ \|\nabla_\gamma J_{O_t}(\gamma) - \nabla_\zeta J_{O_t}(\zeta) \|_2 }{ \|\gamma - \zeta \|_2 } < \infty ;$
\item[(P2)] $ \sup_{\gamma \neq \zeta} \frac{ \|\nabla_\gamma^2 J_{O_t}(\gamma) - \nabla_\zeta^2 J_{O_t}(\zeta) \|_2 }{ \|\gamma - \zeta \|_2 } < \infty ;$
\item[(P3)] $ \sup_\gamma \| \nabla_\gamma J_{O_t}(\gamma) \|_2 < \infty $,
\end{enumerate}
for both $O_t \in \{S_t,Q_t\}$.

\begin{proof}[Proof of P1-P3]
Denote the additive contribution of a single sample to the loss function as $l_t(x,\gamma) = (f_t(x) - g(x;\gamma))^2$.
First we will show that under \Cref{assum:cv_new1} and \Cref{assum:cv_new2}, we have: for each $t$ and $x \in D_t$, the function $\gamma \mapsto \nabla_\gamma \ell_t(x ; \gamma)$ is bounded and Lipschitz; and for each $t$ and $x \in D_t$, the function $\gamma \mapsto \nabla_\gamma^2 \ell_t(x ; \gamma)$ is Lipschitz. 
Then (P1-P3) follow immediately as $J_{Q_t}(\gamma)= \frac{1}{\vert Q_t \vert}\sum_{x\in Q_t}l_t(x; \gamma)$ and $J_{S_t}(\gamma)= \frac{1}{\vert S_t \vert}\sum_{x\in S_t}l_t(x; \gamma)$. 

From direct calculation, we have:
\begin{align*}
\nabla_\gamma \ell_t(x ; \gamma) & = - 2 (f_t(x) - g(x ; \gamma)) \nabla_\gamma g(x;\gamma) \\
\nabla_\gamma^2 \ell_t(x ; \gamma) & = 2 (f_t(x) - g(x ; \gamma)) \nabla_\gamma g(x;\gamma) \nabla_\gamma g(x;\gamma)^\top  - 2 (f_t(x) - g(x ; \gamma)) \nabla_\gamma^2 g(x;\gamma) \\
& = 2 (f_t(x) - g(x ; \gamma)) \left[ \nabla_\gamma g(x;\gamma) \nabla_\gamma g(x;\gamma)^\top - \nabla_\gamma^2 g(x;\gamma) \right] 
\end{align*}
and taking differences:
\begin{align*}
\| \nabla_\gamma \ell_t(x ; \gamma) - \nabla_\zeta \ell_t(x ; \zeta) \|_2 & = \| - 2 (f_t(x) - g(x ; \gamma)) \nabla_\gamma g(x;\gamma) + 2 (f_t(x) - g(x ; \zeta)) \nabla_\zeta g(x;\zeta) \|_2 \\
& \leq 2 |f_t(x)| \| \nabla_\gamma g(x ; \gamma) - \nabla_\zeta g(x ; \zeta) \|_2 \\
& \qquad + 2 \| g(x ; \gamma) \nabla_\gamma g(x;\gamma) - g(x ; \zeta) \nabla_\zeta g(x;\zeta) \|_2 \\
& \leq 2 |f_t(x)| \| \nabla_\gamma g(x ; \gamma) - \nabla_\zeta g(x ; \zeta) \|_2 \\
& \qquad + 2 | g(x ; \gamma) | \| \nabla_\gamma g(x;\gamma) - \nabla_\zeta g(x ; \zeta) \|_2 + 2 \| \nabla_\zeta g(x;\zeta)\|_2  |g(x;\gamma) - g(x;\zeta)| .
\end{align*}
So, for each $t$ and $x \in D_t$, the function $\gamma \mapsto \nabla_\gamma \ell_t(x ; \gamma)$ is bounded and Lipschitz when the functions $\gamma \mapsto g(x;\gamma)$ and $\gamma \mapsto \nabla_\gamma g(x;\gamma)$ are bounded and Lipschitz (i.e. \Cref{assum:cv_new1}).

Then taking differences and bounding terms in a similar manner, we have, 
\begin{align*}
\| \nabla_\gamma^2 \ell_t(x ; \gamma) - \nabla_\zeta^2 \ell_t(x ; \zeta) \|_2 & \leq 2 |f_t(x)| \| \nabla_\gamma g(x;\gamma) \nabla_\gamma g(x;\gamma)^\top - \nabla_\gamma^2 g(x;\gamma) \\
& \hspace{60pt} - \nabla_\zeta g(x;\zeta) \nabla_\zeta g(x;\zeta)^\top + \nabla_\zeta^2 g(x;\zeta) \|_2 \\
& \qquad + 2|g(x;\gamma)| \| \nabla_\gamma g(x;\gamma) \nabla_\gamma g(x;\gamma)^\top - \nabla_\gamma^2 g(x;\gamma) \\
& \hspace{100pt} - \nabla_\zeta g(x;\zeta) \nabla_\zeta g(x;\zeta)^\top + \nabla_\zeta^2 g(x;\zeta) \|_2 \\
& \qquad + 2\|  \nabla_\zeta g(x;\zeta) \nabla_\zeta g(x;\zeta)^\top - \nabla_\zeta^2 g(x;\zeta) \|_2 |g(x;\gamma) - g(x;\zeta)|
\end{align*}
So for each $t$ and $x \in D_t$, the function $\gamma \mapsto \nabla_\gamma^2 \ell_t(x ; \gamma)$ is Lipschitz when the functions $\gamma \mapsto \nabla_\gamma g(x;\gamma) \nabla_\gamma g(x;\gamma)^\top - \nabla_\gamma^2 g(x;\gamma)$ are bounded and Lipschitz (i.e. \Cref{assum:cv_new2}).
\end{proof}

\paragraph{Proof of \Cref{theoCV:cvepsionexits}:} 
\begin{proof}
    Assumption (A1) is automatically satisfied.  (P1) and (P2) above imply (A2) and (A3). (P3) above implies (A4). 
    
    Note that \Cref{assum:cv_new1} implies (A5). This is because, for each $t$, $x\in D_t$, we have $\sup_\gamma l_t(x; \gamma):=\sup_\gamma(f_t(x) - g(x;\gamma))^2 < \infty$ as we assume that $\gamma \mapsto g(x;\gamma)$ is bounded and $f_t(x)$ is constant in $\gamma$. Thus, $\sup_\gamma J_{O_t}(\gamma) =\frac{1}{|O_t|} \sum_{x \in O_t} l_t(x;\gamma)<\infty$ where $O_t$ can be either $S_t$ or $Q_t$. So $\sup_\gamma \|J_{S_t}(\gamma) - J_{Q_t}(\gamma)\|_2 <\infty$.
    
    Then, \Cref{theoCV:cvepsionexits} follow from the conclusion of \Cref{theo:theo9_coro10_in_conver_gbml}.
\end{proof}

\subsection{Proof of Corollary~\ref{theoCV:from_r_to_ri}}
\label{appdx:proof_thereom2}

\begin{proof} Since \Cref{assum:cv_new1} and \Cref{assum:cv_new2} imply (A1) to (A5) in \Cref{appdx:previous_results}, we will use the constants defined earlier in \Cref{appdx:previous_results} here as well.
Firstly, note that given $\hat{\gamma}_{\epsilon}$, with 
\begin{talign*}
    \alpha < \frac{\exp(\frac{\log 2}{2 L})-1}{\chi} = \frac{2^{\frac{1}{2L}}-1}{\chi} ,
\end{talign*}
we have:
$\E\| \E_{t} [ \nabla J_{Q_{t}}(\hat{\gamma}_{\epsilon}) ] \|_2 \leq\frac{1}{C_1'}\epsilon + \frac{C_2'}{C_1'}$ by taking $\gamma = \hat{\gamma}_{\epsilon}$ in \Cref{lemma:lemma19_in_conver_gbml}.

If then additionally $ \nabla^2 J_{Q_t}(\gamma) \succeq \mu I_{p+1}$ holds, by (9.11) in \cite{boyd2004convex} we have,
\begin{talign*}
    \|\gamma -\gamma_t^* \|_2 \leq \frac{2}{\mu} \|\nabla 
    J_{Q_t}(\gamma)\|_2.
\end{talign*}

Taking the expectation of both sides, we then have
\begin{talign*}
         \E_{t} [ \|\gamma - \gamma_t^*\|_2 ] &\leq \frac{2}{\mu} \E_{t}[\|\nabla J_{Q_t}(\gamma)\|_2] \\
         &\overset{(i)} \leq \frac{2}{\mu} (\|\E_{t }[\nabla J_{Q_t}(\gamma)] \|_2 +\sigma ) ,
\end{talign*}
where $(i)$ follows from \citep{ji2022theoretical_ms_gbml} (Page 35, Line 8). Take $\gamma = \hat{\gamma}_{\epsilon}$ and take the expectation of both sides. Then by \Cref{theoCV:cvepsionexits},
\begin{talign*}
     \E[\E_{t} [\|\hat{\gamma}_{\epsilon} - \gamma_t^*\|_2 ]] 
     &\leq \frac{2}{\mu} \E [ \|\E_{t}[\nabla J_{Q_t}(\hat{\gamma}_{\epsilon})] \|_2 ] + \frac{2\sigma }{\mu} \\
     &\leq \frac{2}{\mu} \left(\frac{1}{C_1'}\epsilon + \frac{C_2'}{C_1'}\right)+\frac{2\sigma }{\mu} \\
     &= \frac{2}{\mu C_1'}\epsilon + \frac{2(\sigma C_1' + C_2')}{\mu C_1'} \\
     &=\frac{C_{1}}{\mu}\epsilon + \frac{C_{2}}{\mu} ,
\end{talign*}
where $C_{1} = \frac{2}{C_1'}$ and $C_{2} =\frac{2(\sigma C_1' + C_2')}{C_1'} $.
\end{proof}

\section{Experimental Details}
\label{appdx:experimental_details}

In this section, we provide more experimental details and implementation details of Neural-CVs and Meta-CVs. Details of the synthetic example are presented in \Cref{appdx:experiments_oscillatory}. Details of the boundary-value ODE are provided in \Cref{appdx:boundary_value_ODEs}. Details of Bayesian inference for the Lotka--Volterra system are provided in \Cref{appdx:experiments_lotka}. Details of the Sarcos robot arm are presented in \Cref{appdx:sarcos}.

\subsection{Experiment: Oscillatory Family of Functions}
\label{appdx:experiments_oscillatory}

Our environment $\rho$ consists of independent distributions on each element of $a$. For $a_1$, we select a $\textsf{Unif}(0.4, 0.6)$, whilst for all other parameters we select a $\textsf{Unif}(4,6)$. Each task is of the form $\mathcal{T}_t = \{f_t(x;a_t), \pi_t\}$ where $a_t := (a_{t,1}, a_{t,2:d+1})^\top$ is a sample from $\rho$. This creates potentially infinite number of integral estimation tasks as $a$ is continuous. The target distributions are $\pi_1(x) = \ldots =\pi_T(x) = \textsf{Unif}(0,1)^d$ where $d$ is the dimension of $x$.

For all experiments of this example, we set the neural network identical for both Meta CVs and Neural CVs. That is, a fully connected neural network with two hidden layers. Each layer has $80$ neurons while the output layer has $1$ neurons (the output then is multiplied by a identity matrix $I_d$ to used as $\tilde{u}$ where $d$ is the dimension of the input $x$). The total number of parameters of the neural network $p = 80d + 6641$ where $d$ the dimension of the input $x$. The activation function is the sigmoid function. The neural network is served as $\tilde{u}$ and we apply Langevin Stein operator onto $\tilde{u}(x)\delta(x)$ where $\delta(x) = \prod_{j=1}^d x_j(1-x_j)$ to satisfy assumptions in \citep{oates2019convergence}. For experiments in this example, we use Adam as the $\textsc{Update}$ rule in this example and the penalty constant $\lambda$ is set to be $5\times 10^{-6}$.

\paragraph{2-dimensional Oscillatory Family of Functions}
\begin{itemize}
    \item For Meta-CVs: The inner step size $\alpha = 0.01$. The number of inner gradient steps is $L=1$. The meta step size $\eta = 0.002$ for all meta iterations. The number of meta iteration $\nepoch$ is set to be $4,000$. The meta batch size of tasks $B$ is set to be $5$.
    \item  For Neural-CVs: The step size (learning rate) is $0.002$. The number of training epochs for each task is set to be $20$ with batch size $5$.
    
   \item For Control functionals: we use radius basis function $k(x,x') =  \exp( - \frac{\|x-x'\|_2^2}{2v})$ with kernel hyperparameter $v >0$ as the base kernel for control functionals. The hyper-parameter $v$ is tuned by maximising the marginal likelihood of the Stein kernel on $S_t$ for each task. Optimal control functionals are selected by using $S_t$ and then unbiased control functional estimators are constructed by using $Q_t$ of each task.

\end{itemize}

\paragraph{Impact of the Number of Inner Updates $L$}
\begin{itemize}
    \item  For Meta-CVs: The inner step size $\alpha = \frac{0.01}{50 \times L}$ for $L \in \{1,3,5,7,10\}$. The meta step size $\eta = 0.002$ for all meta iterations. The number of meta iteration $\nepoch$ is set to be $4,000$. The meta batch size of tasks $B$ is set to be $5$.
\end{itemize}

\paragraph{Impact of Dimensions}
\begin{itemize}
    \item For Meta-CVs: The inner step size $\alpha = 0.01$. The number of inner gradient steps is $L=1$. The meta step size $\eta = 0.002$ for all meta iterations. The number of meta iteration $\nepoch$ is set to be $4,000$. The meta batch size of tasks $B$ is set to be $5$.
    \item  For Neural-CVs: The step size (learning rate) is $0.002$. The number of training epochs for each task is set to be $20$ with batch size $5$.

   \item For Control functionals: we use radius basis function $k(x,x') =  \exp(-\frac{\|x-x'\|_2^2}{2v})$ with kernel hyperparameter $v >0$ as the base kernel for control functionals. The hyper-parameter $v$ is tuned by maximising the marginal likelihood of the Stein kernel on $S_t$ for each task. Optimal control functionals are selected by using $S_t$ and then unbiased control functional estimators are constructed by using $Q_t$ of each task.

\end{itemize}

\paragraph{Impact of $B$ and $\nepoch$ of Meta-CVs}
\begin{itemize}
    \item The inner step size $\alpha =0.01$. The number of inner gradient steps is $L=1$. The meta step size is $\eta= 0.002$ for all meta iterations.
\end{itemize}

\subsection{Experiment: Boundary Value ODEs}
\label{appdx:boundary_value_ODEs}
For all experiments of this example, we set the neural network identical for both Meta-CVs and Neural-CVs. That is, a fully connected neural network with three hidden layers. Each layer has $80$ neurons while the output layer has $1$ neurons. The total number of parameters of the neural network $p =13,201$. The activation function is the sigmoid function. We use Adam as the $\textsc{Update}$ rule in this example and the penalty constant $\lambda$ is set to be $5\times 10^{-6}$.

\begin{itemize}
    \item For Meta-CVs: The inner step size $\alpha = 0.01$ and the meta step size $\eta = 0.002$ for all meta iterations. The number of inner updates is $L=1$. The number of meta iteration $\nepoch$ is set to be $2,000$. The meta batch size of tasks is set to be $5$. 
    \item For Neural-CVs: The step size (learning rate) is $0.002$. The number of training epochs for each task is set to be $20$ with batch size $5$.
\end{itemize}

\subsection{Experiment: Bayesian Inference of Lotka-Volterra System}
\label{appdx:experiments_lotka}

The $\log$-$\exp$ transform is used on the model parameters $x$ to avoid constrained parameters on the ODE directly. We reparameterised the Lotka—Volterra system as,
\begin{talign*}
    \frac{\mathrm{d}u_1(s)}{\mathrm{d}s} &= \tilde{x}_1 u_1(s)- \tilde{x}_2 u_1(s) u_2(s)\\
    \frac{\mathrm{d}u_2(s)}{\mathrm{d}s} &= \tilde{x}_3 u_1(s) u_2(s)- \tilde{x}_4 u_2(s),
\end{talign*}
where 
\begin{talign*}
    &\tilde{x}_1 =\exp(x_1), \tilde{x}_2  = \exp(x_2) , \\
    &\tilde{x}_3 = \exp(x_3),  \tilde{x}_4 = \exp(x_4),
\end{talign*}
where $u_1$ and $u_2$ represents the number of preys and predators, respectively.

\noindent The model is,
\begin{talign*}
    y_{1}(0) \sim \text{Log-Normal}(\log \tilde{x}_5, \tilde{x}_7)\\
    y_{2}(0) \sim \text{Log-Normal}(\log \tilde{x}_6, \tilde{x}_8) \\
    y_{1}(s) \sim \text{Log-Normal}(\log u_1(s), \tilde{x}_7) \\
    y_{2}(s) \sim \text{Log-Normal}(\log u_2(s), \tilde{x}_8) 
\end{talign*}
where 
\begin{talign*}
    &\tilde{x}_5: = \exp(x_5), \tilde{x}_6: = \exp(x_6)\\
    &\tilde{x}_{7}:=\exp(x_7), \tilde{x}_{8}=\exp(x_8). 
\end{talign*}

\noindent By doing so, $x$ is then on the whole $\R^8$. As a result, the prior distribution $\pi(x)$ is defined on $\R^8$ and Stan will return the scores of these parameters directly as these 8 parameters $x$ themselves are unconstrained through manually reparameterisation directly.

\noindent Priors are,
\begin{talign*}
  x_1 ,x_4 &\sim \text{Normal}(0, 0.5^2) \\
  x_2 , x_3 &\sim \text{Normal}(-3, 0.5^2) \\
  x_5, x_6 &\sim \text{Normal}(\log 10,1^2) \\
  x_7, x_8 &\sim \text{Normal}(-1,1^2)
\end{talign*}

\paragraph{Inference of $x_1$ and $x_2$} 
\begin{itemize}
    \item For both Meta-CVs and Neural-CVs: We use a fully connected neural network with $3$ hidden layers. Each layer has $5$ neurons while the output layer has $8$ neurons. The total number of parameters of the neural network $p =153$.  The activation function is the tanh function. All parameters of neural networks are initialised with a Gaussian distribution with zero mean and standard deviation $0.01$ except of $\gamma_{t,0}$ is initialised at the Monte Carlo estimator of each task. We use Adam as the $\textsc{Update}$ rule in this example and the penalty constant $\lambda$ is set to be $5\times 10^{-5}$.
    
    \item For Meta-CVs: The inner step size $\alpha = 0.0001$. The number of inner gradient steps is $L=1$. The meta step size was initialised at $0.001$ with a step size decay ($\eta_{i+10} = 0.9 \eta_{i} $) every $10$ meta iterations. The number of meta iteration $\nepoch$ is set to be $2,000$. The meta batch size of tasks $B$ is set to be $5$. We only use 100 tasks (sub-populations) for learning the Meta-CVs. For each of these 100 tasks, we have more than $N_t$ data points (also because MCMC sampler will return more than $N_t$ samples, so we reuse all of them) such that we can learn Meta-CV with $\nepoch = 2000$ and $B=5$.
    
    \item For Neural-CVs: The step size (learning rate) is $0.001$. The number of training epochs for each task is set to be $20$ with batch size $5$.
\end{itemize}

\paragraph{Inference of $x_3$ and $x_4$} 
\begin{itemize}
    \item For both Meta-CVs and Neural-CVs: We use a fully connected neural network with $3$ hidden layers. Each layer has $3$ neurons while the output layer has $8$ neurons. The total number of parameters of the neural network $p =83$.  The activation function is the tanh function. All parameters of neural networks are initialised with a Gaussian distribution with zero mean and standard deviation $0.01$ except of $\gamma_{t,0}$ is initialised at the Monte Carlo estimator of each task. We use Adam as the $\textsc{Update}$ rule in this example and the penalty constant $\lambda$ is set to be $5\times 10^{-5}$.
    
    \item For Meta-CVs:  The inner step size $\alpha = 0.001$. The number of inner gradient steps is $L=1$. The meta step size was initialised at $0.001$ with a step size decay ($\eta_{i+10} = 0.9 \eta_{i} $) every $10$ meta iterations. The number of meta iteration $\nepoch$ is set to be $2,000$. The meta batch size of tasks $B$ is set to be $5$. We only use 100 tasks (sub-populations) for learning the Meta-CVs. For each of these 100 tasks, we have more than $N_t$ data points (also because MCMC sampler will return more than $N_t$ samples, so we reuse all of them) such that we can learn Meta-CV with $\nepoch = 2000$ and $B=5$.
    
    \item For Neural-CVs: The step size (learning rate) is $0.001$. The number of training epochs for each task is set to be $20$ with batch size $5$.
\end{itemize}

\subsection{Experiment: Sarcos Robot Arm}
\label{appdx:sarcos}

\paragraph{Approximate Inference of Full Bayesian Gaussian Process Regression} We learn full Bayesian hierarchical Gaussian processes by variational inference \citep{kucukelbir2017automatic_VL, lalchand2020approximate_FB_GP}.

\begin{figure}[ht]
    \centering
   \includegraphics[width=0.5\columnwidth]{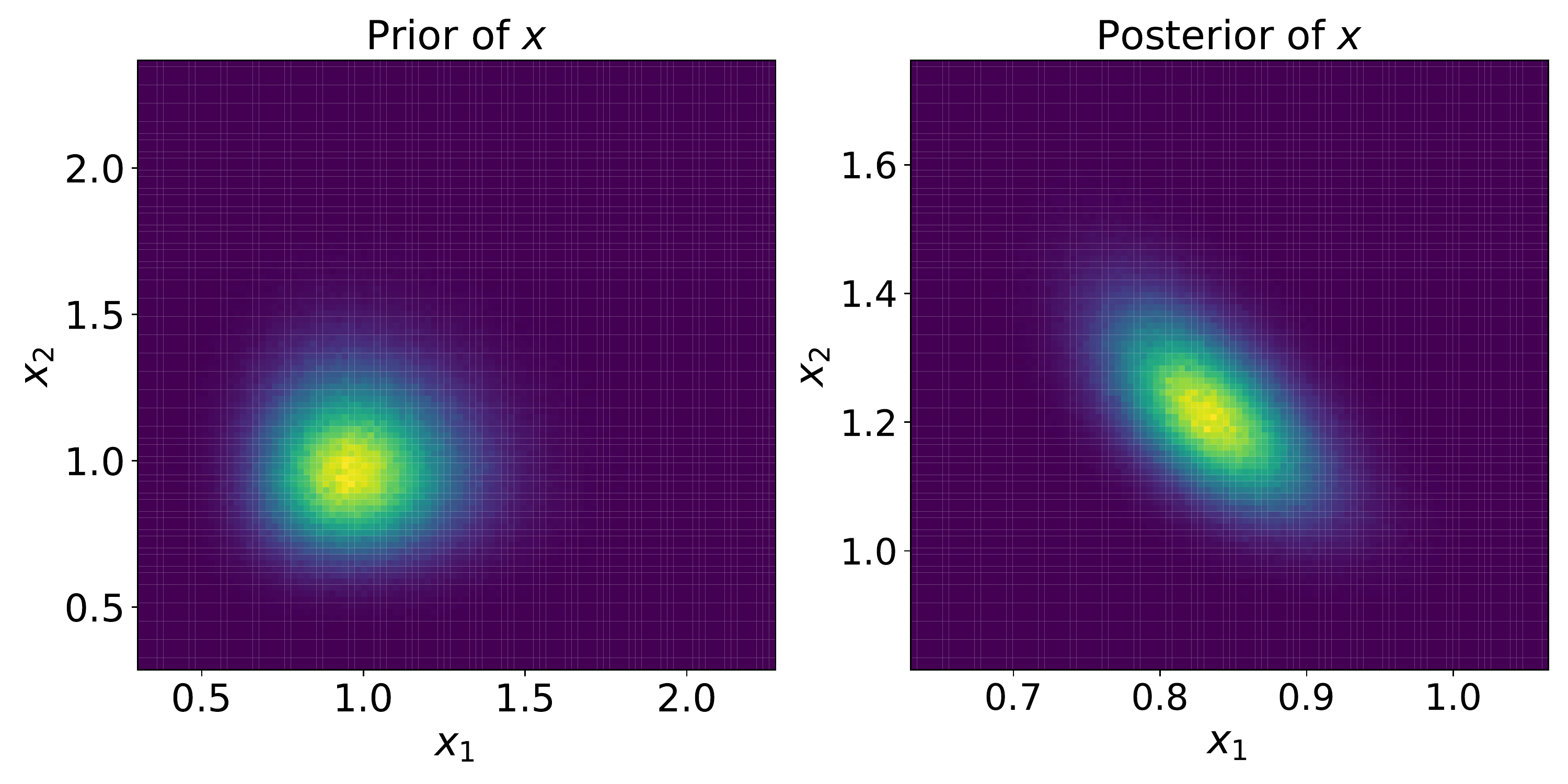}
    \caption{Priors and Posteriors of Kernel Hyper-parameters $x$.}
\label{fig:prior_post_kernelparams}
\end{figure}

We set $\sigma=0.1$, $\pi(x_1) = \textsf{Gamma}(25,25)$ and $\pi(x_2)= \textsf{Gamma}(25,25)$, which is the prior used in \citep{oates2017_CF_for_MonteCarloIntegration}.
We transform the kernel hyper-parameters $x \in \R^{2+}$ to $\eta = g(x)= \log x$ such that we can learn a variational distribution $q_{\phi}(\eta)$ of $\eta$ in $\R^2$ and then transform back to $q(x)$. We use full rank approximation which means the variational family takes the following form:
\begin{talign*}
    q_{\phi}(\eta) = \textsf{N}(\mu, VV^\top) ,
\end{talign*}
with variational parameter $\phi:= \{\mu, V\} \in \R^{p+p(p+1)/2}$ where $\mu$ is a column vector and $V$ is a lower triangular matrix. The objective of variational inference is to maximize the evidence lower with respect to $\phi$, which is given by,
\begin{talign*}
    \textsf{ELBO}(\phi) &= \E_{q_\phi} [\log p(y_{1:q}, e^{\eta}) + \log \vert \textsf{Jacobian}_{g^{-1}}(\eta) \vert] - \E_{q_{\phi}}[\log q_{\phi}(\eta)] \\
    &=  \E_{q_\phi} [\log p(y_{1:q}|e^{\eta}) +\log \pi(e^{\eta}) + \log \vert \textsf{Jacobian}_{g^{-1}}(\eta) \vert] - \E_{q_{\phi}}[\log q_{\phi}(\eta)]
\end{talign*}
The expectations involved in $\textsf{ELBO}(\phi)$ are approximated by Monte Carlo estimators and we use re-parametrization trick \citep{kingma2013_VAE} to learn $\phi$. \Cref{fig:prior_post_kernelparams} demonstrates the prior and the corresponding posterior of the kernel hyper-parameters $x=(x_1, x_2)$ (in the form of $2$d histograms).

\paragraph{Settings}
\begin{itemize}
   \item For both Meta-CVs and Neural-CVs, a fully connected neural network with $5$ hidden layers. Each layer has $20$ neurons while the output layer has $2$ neurons (the output then is timed by a identity matrix $I_2$ to used as $u$ since $2$ is the dimension of the input $x$). The total number of parameters of the neural network $p =10,401$. The activation function is the sigmoid function.  All parameters of neural networks are initialised with a Gaussian distribution with zero mean and standard deviation $0.001$. We use Adam as the $\textsc{Update}$ rule in this example and the penalty constant $\lambda$ is set to be $1\times 10^{-10}$.
    \item For Meta-CVs: The inner step size $\alpha = 0.01$. The meta step size was initialised at $0.001$ with a step size decay ($\eta_{i+10} = 0.9 \eta_{i} $) every $10$ meta iterations. The number of meta iteration $\nepoch$ is set to be $1,000$. The meta batch size of tasks $B$ is set to be $1$. 
    \item For Neural CV: The step size (learning rate) is $0.001$. The number of training epochs for each task is set to be $20$ with batch size $5$.
   \item For Control functionals: we use radius basis function $k(x,x') =  \exp(-\frac{\|x-x'\|_2^2}{2v})$ with kernel hyperparameter $v >0$ as the base kernel for control functionals. The hyper-parameter $v$ is tuned by maximising the marginal likelihood with the Stein kernel on $S_t$ for each task. Optimal control functionals are selected by using $S_t$ and then unbiased control functional estimators are constructed by using $Q_t$ of each task.

\end{itemize}

\begin{figure}[ht]
    \centering
   \includegraphics[width=0.5\columnwidth]{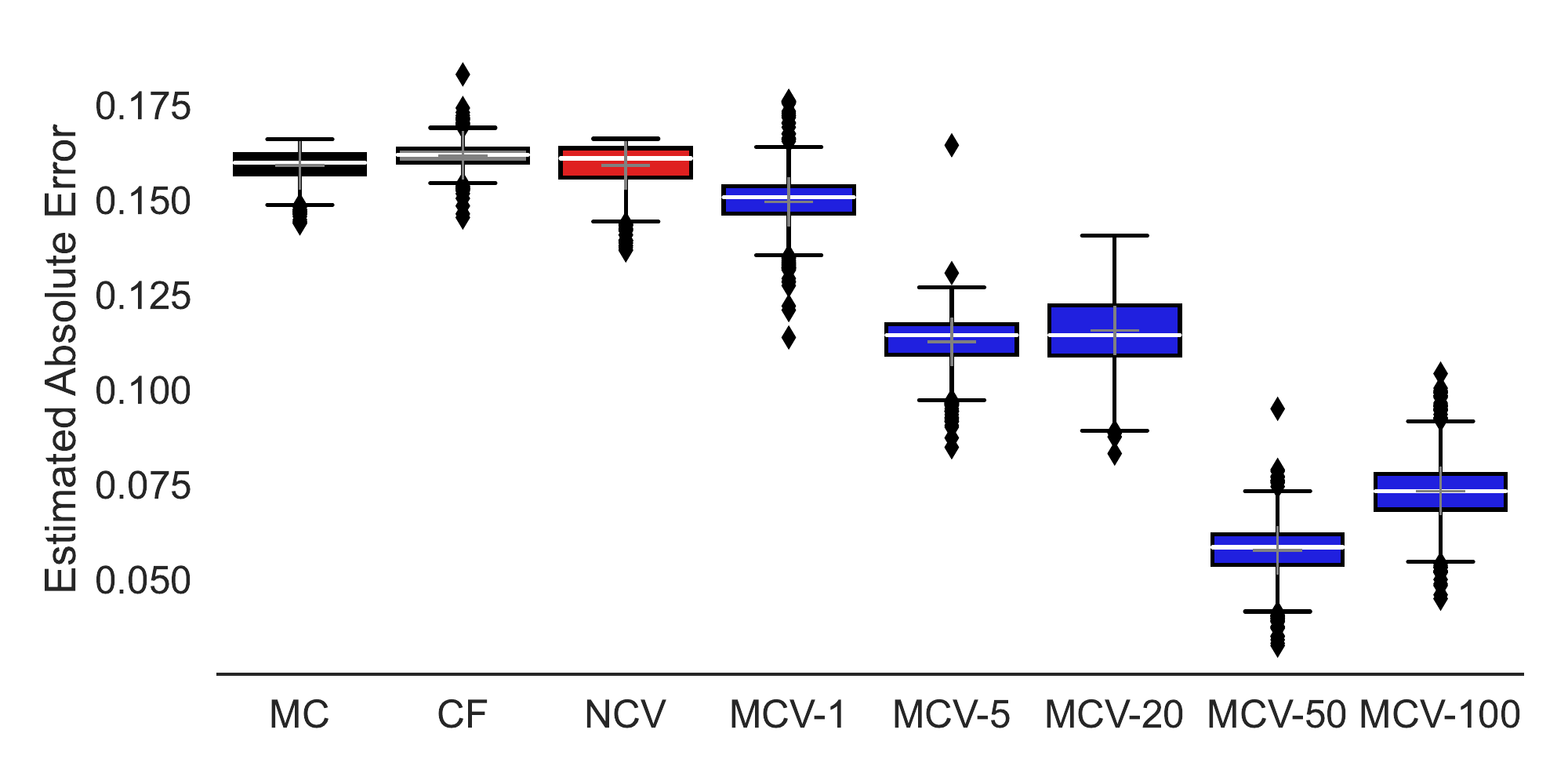}
    \caption{Estimated absolute errors over the same training states (which are used for learning the Meta-CV) of the Sarcos anthropomorphic robot arm (CF: Control functionals; NCV: Neural-CVs; MCV-L: Meta-CVs with L inner steps).}
\label{fig:fb_Sarcos_gammaprior_investinsteps_sameMetaTrainTest}
\end{figure}

\paragraph{Extra Experiments} In addition, we test the performance of Meta-CVs on the same tasks used for learning the Meta-CV. Under the same setting described above, the comparisons between Meta-CVs and other methods are presented in \Cref{fig:fb_Sarcos_gammaprior_investinsteps_sameMetaTrainTest}.

\end{document}